\documentclass[floatfix, preprint, onecolumn, prd,showpacs, showkeys, preprintnumbers, nofootinbib, superscriptaddress]{revtex4-1}
\usepackage[sort&compress]{natbib}
\usepackage[utf8]{inputenc}
\usepackage{times}
\usepackage{amssymb,amsbsy,amsmath,amsfonts}
\usepackage{graphicx}
\usepackage{float}
\usepackage{bm}
\usepackage{color}
\usepackage{morefloats}
\usepackage{rotating}
\usepackage{srcltx}
\usepackage{slashed}
\usepackage{subfigure}
\usepackage{multirow}
\usepackage{verbatim}
\usepackage{hyperref}
\usepackage{tabularx}

\usepackage{color}

\begin{document}
\title{Renormalizability of leading order covariant chiral nucleon-nucleon interaction}

\author{Chun-Xuan Wang}
\affiliation{School of Physics,  Beihang University, Beijing 102206, China}

\author{Li-Sheng Geng}
\email[E-mail: ]{lisheng.geng@buaa.edu.cn}
\affiliation{School of
Physics,  Beihang University, Beijing 102206, China}
\affiliation{Beijing Key Laboratory of Advanced Nuclear Materials and Physics, Beihang University, Beijing 100191, China }\affiliation{Beijing  Advanced Innovation Center for Big Data-based Precision Medicine, Beihang University, Beijing 100191, China}
\affiliation{School of Physics and Microelectronics, Zhengzhou University, Zhengzhou, Henan 450001, China }

\author{Bingwei Long}
\email{bingwei@scu.edu.cn}
\affiliation{Center for Theoretical Physics, Department of Physics, Sichuan University,  Sichuan 610064, China}

\begin{abstract}
In this work, we study the renormalization group invariance (RGI) of the recently proposed covariant power counting (PC) scheme in the case of nucleon-nucleon scattering [Chin.Phys. C42 (2018) 014103] at leading order (LO). We show that unlike the LO Weinberg case, RGI is satisfied in the $^3P_{0}$ channel, because a term of $pp'$ appears naturally in  the covariant PC scheme at LO.  Another interesting feature  is that the $^1S_{0}$ and $^3P_{1}$ channels are correlated. Fixing the two relevant low energy constants by fitting to the $^1S_{0}$ phase shifts at $T_\mathrm{lab.}=10$ and $25$ MeV with a cutoff $\Lambda$ of $400-650$ MeV, the $^3P_{1}$ phase shifts can be described relatively well. In the limit of $\Lambda\rightarrow \infty$, the $^1S_0$ channel becomes cutoff independent, while RGI is lost in the $^3P_{1}$ channel, consistent with the Wigner bound and the previous observation that the $^{3}P_1$ channel better be
treated perturbatively. As for the $^1P_{1}$ and $^3S_{1}$-$^3D_{1}$ channels, RGI is satisfied,  similar to the Weinberg approach.

\end{abstract}


\maketitle
\section{Introduction}

 Since the pioneering works of Weinberg~\cite{Weinberg:1990rz,Weinberg:1991um}, chiral effective field theory (ChEFT) has been successfully applied to describe the nucleon-nucleon interaction. Today, high precision chiral nuclear forces have become the de facto standard in ab initio nuclear structure and reaction studies~\cite{Epelbaum:2008ga,Machleidt:2011zz,Hammer:2019poc}. Nevertheless, there are still a few outstanding issues in current chiral nuclear forces. One hotly discussed issue is  theri renormalizatheirbility, see, e.g.,  Refs.~\cite{Kaplan:1998tg,Yang:2019hkn}.

  ChEFT is based on the chiral symmetry of QCD and its explicit and spontaneous breaking~\cite{Weinberg:1978kz,Gasser:1983yg,Ecker:1989yg,Ecker:1994gg,Pich:1995bw,Bernard:1995dp}. In ChEFT,
 the long range interaction is provided by the exchange of the Goldstone bosons (pions in the $u$ and $d$ two flavor sector and the pseudoscalar nonet in the $u$, $d$, and $s$ three flavor sector), and the short range interaction is
 described by the so-called low-energy constants (LECs) that encode the effects of degrees of freedom with energies larger than the chiral symmetry breaking scale, $\Lambda_{\chi SB}$. In principle, these LECs can be calculated in the underlying theory, QCD, but in practice they can only be determined  by fitting  either to experimental or lattice QCD data, because of the non-perturbative nature of low energy strong interactions.
 As a result, the predictive power of ChEFT relies on the fact that at a certain order and to a specific observable
 only a finite number of LECs contribute.

 For an EFT,  a proper power-counting (PC) scheme is the most important ingredient in order to perform any calculation.  Current high precision chiral nuclear forces are based on the Weinberg PC (WPC) , or  the so-called naive dimensional analysis (NDA)~\cite{Weinberg:1990rz,Weinberg:1991um}.
However, in the past two decades, one realized that the WPC is inconsistent in the sense that  the so-constructed chiral nuclear force is not renormalization group invariant (RGI), or naively, is not cutoff independent.  Since in any EFT, a separation (cutoff) between high- and low-energy physics should be offset with the LECs once one refits them for  each new cutoff. In the WPC, the inconsistency problem already appears at leading order~\cite{Beane:2000wh,Nogga:2005hy}.

The pursuit of a consistent PC has continued for almost two decades. Using RGI as a guideline, it has been proposed that one can promote some of the high-order terms in the WPC to make
the chiral nuclear force renormalization group invariant at a specific order~\cite{Long:2007vp,Long:2011qx,Long:2011xw,Long:2012ve}. In principle, one can count how many counter terms are needed before calculations are done by solving the Wilson  RG-equation~\cite{Birse:2005um,Barford:2002je,Valderrama:2014vra}. A modified Weinberg approach with Lorentz invariant contact interactions was proposed for nucleon-nucleon scattering in Ref.~\cite{Epelbaum:2012ua} and later applied to study hyperon-nucleon scattering in Ref.~\cite{Li:2016paq}. The modified Weinberg approach was further refined and applied to study baryon-baryon scattering~\cite{Baru:2019ndr,Ren:2019qow} with a different treatment of the one-pion(meson) exchange.

Recently, a covariant power-counting approach for NN scattering was proposed  in Ref.~\cite{Ren:2016jna} with the full structure of the Dirac spinor retained.  At leading order, it already provided a reasonably good description of the phase shifts of angular momentum $J=0$ and $1$ by solving the Kadyshevsky equation~\cite{Kadyshevsky:1967rs}.~\footnote{The numerical results remain almost the same if the Blankenbecler-Sugar~\cite{Blankenbecler:1965gx} equation was used instead.} This framework has also been successfully applied to study hyperon-nucleon interactions~\cite{Li:2016mln,Song:2018qqm,Li:2018tbt}. In Ref.~\cite{Ren:2017yvw}, it was shown that this formulation also provides a good description of the unique features of the $^1S_0$ channel at leading order, in particular the pole position of the virtual bound state and the zero amplitude at the scattering momentum--340 MeV.  According to Ref.~\cite{SanchezSanchez:2017tws}, a proper description of the unique features of the $^1S_0$ channel at leading order can serve as a nontrivial check on the self-consistency of any EFT for the NN interaction. In the present work, in the spirit of Ref.~\cite{Nogga:2005hy}, we study the cutoff dependence of the leading order covariant chiral nucleon-nucleon interaction.

This article is organized as follows. In Sec. II, we briefly introduce the covariant nucleon-nucleon potential. In Sec. III, we study the
cutoff dependence of the partial wave phase shifts of $J=0$ and 1, followed by a short summary in Sec. IV.

\section{THEORETICAL FRAMEWORK}
In Refs.~\cite{Ren:2016jna,Li:2016mln}, similar to the extended-on-mass-shell scheme in the
one-baryon sector~\cite{Gegelia:1999gf,Fuchs:2003qc,Geng:2013xn}, a covariant power counting scheme for the two-baryon sector  was
introduced.  For the nucleon-nucleon interaction, at leading order it contains five covariant four-fermion contact terms without derivatives and the one-pion-exchange (OPE) term~\cite{Ren:2016jna},
\begin{equation}\begin{split}
V_{\textrm{LO}}=V_{\textrm{CTP}}+V_{\textrm{OPE}}.
\end{split}\end{equation}
The contact potential in momentum space reads
\begin{equation}\begin{split}
V_{\textrm{CTP}}&=C_{S}(\overline{u}(\bm{p}',s'_{1})u(\bm{p},s_{1}))(\overline{u}(-\bm{p'},s'_{2})u(-\bm{p},s_{2}))\\
&+C_{A}(\overline{u}(\bm{p}',s'_{1})\gamma_{5}u(\bm{p},s_{1}))(\overline{u}(-\bm{p'},s'_{2})\gamma_{5}u(-\bm{p},s_{2}))\\
&+C_{V}(\overline{u}(\bm{p}',s'_{1})\gamma_{\mu}u(\bm{p},s_{1}))(\overline{u}(-\bm{p'},s'_{2})\gamma_{\mu}u(-\bm{p},s_{2}))\\
&+C_{AV}(\overline{u}(\bm{p}',s'_{1})\gamma_{\mu}\gamma_{5}u(\bm{p},s_{1}))(\overline{u}(-\bm{p'},s'_{2})\gamma_{\mu}\gamma_{5}u(-\bm{p},s_{2}))\\
&+C_{T}(\overline{u}(\bm{p}',s'_{1})\sigma_{\mu\nu}u(\bm{p},s_{1}))(\overline{u}(-\bm{p'},s'_{2})\sigma^{\mu\nu}u(-\bm{p},s_{2})),
\end{split}\end{equation}
where $C_{S,A,V,AV,T}$ are the LECs and $u(\bar{u})$ are the Dirac spinors,
\begin{equation}\begin{split}
u(\bm{p},s)=N_{p}\begin{pmatrix} 1\\ \frac{\bm{\sigma}\cdot \bm{p}}{E_{p}+M} \end{pmatrix}\chi_{s},\quad N_{p}=\sqrt{\frac{E_{p}+M}{2M}}
\end{split}\end{equation}
with  the Pauli spinor $\chi_s$ and $E_p$ ($M$) the nucleon energy (mass). The one-pion-exchange potential in momentum space is
\begin{equation}\begin{split}
V_{\textrm{OPE}}(\bm{p}',\bm{p})=-\frac{g^{2}_{A}}{4f^{2}_{\pi}}\frac{(\overline{u}(\bm{p}',s'_{1})\bm{\tau_1}\gamma^{\mu}\gamma_{5}q_{\mu}u(\bm{p},s_{1}))(\overline{u}(-\bm{p'},s'_{2})\bm{\tau_2}\gamma^{\nu}\gamma_{5}q_{\nu}u(-\bm{p},s_{2}))}{(E_{p'}-E_{p})^2-(\bm{p'}-\bm{p})^2-m^{2}_{\pi}}
\end{split}\end{equation}
where $m_{\pi}$ is the pion mass, $\bm{p}$ and $\bm{p'}$ are initial and final three momentum, $g_{A}=1.267$, and $f_{\pi}=92.4 \textrm{MeV}$.
Note that the leading order relativistic potentials already contain all the six spin operators needed to describe nucleon-nucleon scattering.

The contact potentials can be projected into different partial waves in the  $|LSJ\rangle$ basis, which read
\begin{eqnarray}
V_{1S0}&=&\xi_{N}[C_{1S0}(1+R^{2}_{p}R^{2}_{p'})+\hat{C}_{1S0}(R^{2}_{p}+R^{2}_{p'})]\nonumber\\
&=&4\pi C_{1S0}+ \pi (C_{1S0}+\hat{C}_{1S0})(\frac{p^2}{M^2}+\frac{p'^2}{M^2})+\cdots,\label{eq:1s0}\\
V_{3S1}&=&\frac{\xi_{N}}{9}[C_{3S1}(9+R^2_{p}R^2_{p'})+\hat{C}_{3S1}(R^2_{p}+R^2_{p'})]\nonumber\\
&=&4\pi C_{3S1}+\pi(C_{3S1}+ \frac{\hat{C}_{1P1}}{9})(\frac{p^2}{M^2}+\frac{p'^2}{M^2})+\cdots,\\
V_{3D1}&=&\frac{8\xi_{N}}{9}C_{3D1}R^2_{p}R^2_{p'}=\frac{2\pi C_{3D1}}{9M^2}pp',\\
V_{3S1-3D1}&=&\frac{2\sqrt{2}\xi_{N}}{9}(C_{3S1}R^2_{p}R^2_{p'}+\hat{C}_{3S1}R^2_{p})=\frac{2\sqrt{2}}{9}\pi \hat{C}_{3S1}\frac{p^2}{M^2}+\cdots,\\
V_{3D1-3S1}&=&\frac{2\sqrt{2}\xi_{N}}{9}(C_{3S1}R^2_{p}R^2_{p'}+\hat{C}_{3S1}R^2_{p'})=\frac{2\sqrt{2}}{9}\pi \hat{C}_{3S1}\frac{p'^2}{M^2}+\cdots,\\
V_{3P0}&=&-2\xi_{N}C_{3P0}R_{p}R_{p'}=\frac{-2\pi C_{3P0}}{M^2}pp'
,\\
V_{1P1}&=&-\frac{2\xi_{N}}{3}C_{1P1}R_{p}R_{p'}=\frac{-2\pi C_{1P1}}{3M^2}pp',\label{eq:1p1}\\
V_{3P1}&=&-\frac{4\xi_{N}}{3}C_{3P1}R_{p}R_{p'}=\frac{-4\pi C_{3P1}}{3M^2}pp',
\end{eqnarray}
where $\xi_{N}=4\pi N^2_{p}N^2_{p'}, R_{p}=|\bm{p}|/(E_{p}+M)$, $R_{p'}=|\bm{p'}|/(E_{p'}+M)$, $p$ and $p'$ are absolute value of $\bm{p}$ and $\bm{p'}$, and ``$\cdots$'' denote higher order chiral terms in the WPC. Note that the expansion in $1/M$ shown for 
 $V_{1S0}$, $V_{3S1}$, $V_{3S1-3D1}$, and $V_{3D1-3S1}$ are only done to guide the comparison with the Weinberg approach. In our study, we use the full potential without any approximations.
The coefficients in the partial
waves  are linear combination of the LECs appearing in the Lagrangian,
\begin{eqnarray}\label{eq:con}
C_{1S0}&=&(C_{S}+C_{V}+3C_{AV}-6C_{T}),\nonumber\\
\hat{C}_{1S0}&=&(3C_{V}+C_{A}+C_{AV}-6C_{T}),\nonumber\\
C_{3P0}&=&(C_{S}-4C_{V}+C_{A}-4C_{AV}-12C_{T}),\nonumber\\
C_{1P1}&=&(C_{S}+C_{A}+4C_{T}),\nonumber\\
C_{3P1}&=&(C_{S}-2C_{V}-C_{A}+2C_{AV}),\\
C_{3S1}&=&(C_{S}+C_{V}-C_{AV}+2C_{T}),\nonumber\\
\hat{C}_{3S1}&=&3(C_{V}-C_{A}-C_{AV}-2C_{T}),\nonumber\\
C_{3D1}&=&(C_{S}+C_{V}-C_{AV}+2C_{T}).\nonumber
\end{eqnarray}
We note that three of the eight partial wave coefficients are correlated, namely,
\begin{eqnarray}\label{Fitting formula}
C_{3S1}&=&C_{3D1},\nonumber\\
\hat{C}_{1S0}&=&C_{1S0}-C_{3P1},\\
\hat{C}_{3S1}&=&3C_{3S1}-3C_{1P1}\nonumber.
\end{eqnarray}

A few remarks are in order. First, it is clear that in the limit of $M\rightarrow\infty$, only two LECs in the $^1S_0$ and $^3S_1$ channels remain, in agreement with the WPC. Second, the retainment of the full Dirac spinors in the Lagrangian not only leads to additional terms in the  $^1S_0$ and $^3S_1$ partial waves~\footnote{A large contribution of the correction terms is known to
be essential to describe the $^1S_0$ phase shifts~\cite{Soto:2007pg,Long:2013cya}}, but also provides contributions to other channels which are counted as of higher  (than LO) order in the WPC.  These new contributions
will not only affect the description of the covariant nucleon-nucleon phase shifts but also the renormalizability of the chiral nuclear force. The latter is the main focus of the present work.  Third, in the covariant PC, some of the LECs contribute to different partial waves, which is different from the WPC, where a LEC only contributes to a particular partial wave. It should be noted that the above correlations are only valid at leading order, as can be explicitly checked using the higher order Lagrangians constructed in Ref.~\cite{Xiao:2018jot}.

To take into account the non-perturbative nature of the
nucleon-nucleon interaction, we solve the following Kadyshevsky equation with the above-obtained kernel potential, $V_\mathrm{LO}(\bm{p}',\bm{p})$,
\begin{equation}\label{eq:Kady}\begin{split}
T(\bm{p}',\bm{p})=V(\bm{p}',\bm{p})+\int\frac{d^{3}k}{(2\pi)^3}V(\bm{p}',\bm{k})\frac{M^2}{2E^2_{k}}\frac{1}{E_{p}-E_{k}+i\varepsilon}T(\bm{k},\bm{p})
\end{split}\end{equation}
To avoid ultraviolet divergence, we need to introduce a regulator $f(p,p')$. In
principle, physical observables should be independent of the choice of the  regulator if the EFT is properly formulated, i.e., the EFT is RGI. Here we choose the commonly used separable cutoff function in momentum space, $f(p,p')=\textrm{exp}\big{[}\frac{-\bm{p}^{2n}-\bm{p}'^{2n}}{\Lambda^{2n}}\big{]}$ with $n=2$. The convenience of such a regulator lies in that it only depends on initial and final momenta so that it does not mix partial wave decomposition.

\section{RESULTS AND DISCUSSION}
In this section, we present the fitting strategy and results of the RG-analysis in  all the $\textrm{J}=0,1$ partial waves.

\subsection{Fitting strategy}
Numerically, we fit the Nijmegen partial wave phase shifts of the $np$ channel~\cite{Stoks:1993tb}. At  LO, there are five linear independent LECs and they can be divided into three groups according to Eq.(~\ref{Fitting formula}):  1) $C_{3P0}$, 2) $C_{1S0},\hat{C}_{1S0},C_{3P1}$,  and 3) $C_{3S1},\hat{C}_{3S1},C_{1P1}$,
In groups 2 and 3, only two of the three LECs are linear independent.

Since our aim is to study the dependence of observables, here phase shifts, on the chosen value of the cutoff, we fit the coefficients of the partial wave potentials rather than the LECs of the Lagrangian. First, we fit $C_{3P0}$ to the $^3P_{0}$ phase shift at $T_\mathrm{lab.}=10$ MeV. Then we fit $C_{3P1}$ and $C_{1S0}$ to two $^1S_{0}$ phase shifts at $T_\mathrm{lab.}=10$ and 25 MeV. Last, we fit $C_{1P1}$ and $C_{3S1}$ to the $^1P_{1}$ and $^3S_{1}$ phase shifts at $T_\mathrm{lab.}=10$ MeV. In the fitting, we take into account that $\hat{C}_{1S0}=C_{1S0}-C_{3P1}$ and $\hat{C}_{3S1}=3C_{3S1}-3C_{1P1}$.

The reason why we adopt such a fitting strategy is that $^3P_{1}$ is not renormalization group invariant with a potential  of the form $V_{3P1}=\textrm{OPE}+C_{3P1}pp'$, as shown in Ref.~\cite{Yang:2009kx}. Therefore, we better use the two $^1S_{0}$ constants to predict the $^3P_{1}$
phase shifts since they are coupled in the covariant PC scheme. There will be more discussions in Sec.~\ref{sec-1S03P1}.

In the fitting, we define $\tilde{\chi}^2$ as $\tilde{\chi}^2=(\delta_{\textrm{LO}}-\delta_{\textrm{PWA}})^2$,  namely
we neglect the uncertainties of the data as they are much smaller compared
to higher chiral order contributions. In our study the momentum cutoff $\Lambda$ is varied from 0.4 GeV to 10 GeV except for the $^1S_{0}$ and $^3P_{1}$ channels, due to the reasons explained below.

\subsection{One-pion exchange in the covariant approach}
\begin{figure}
\centering
\begin{tabular}{ccc}
{\includegraphics[width=5.5cm,height=3.5cm]{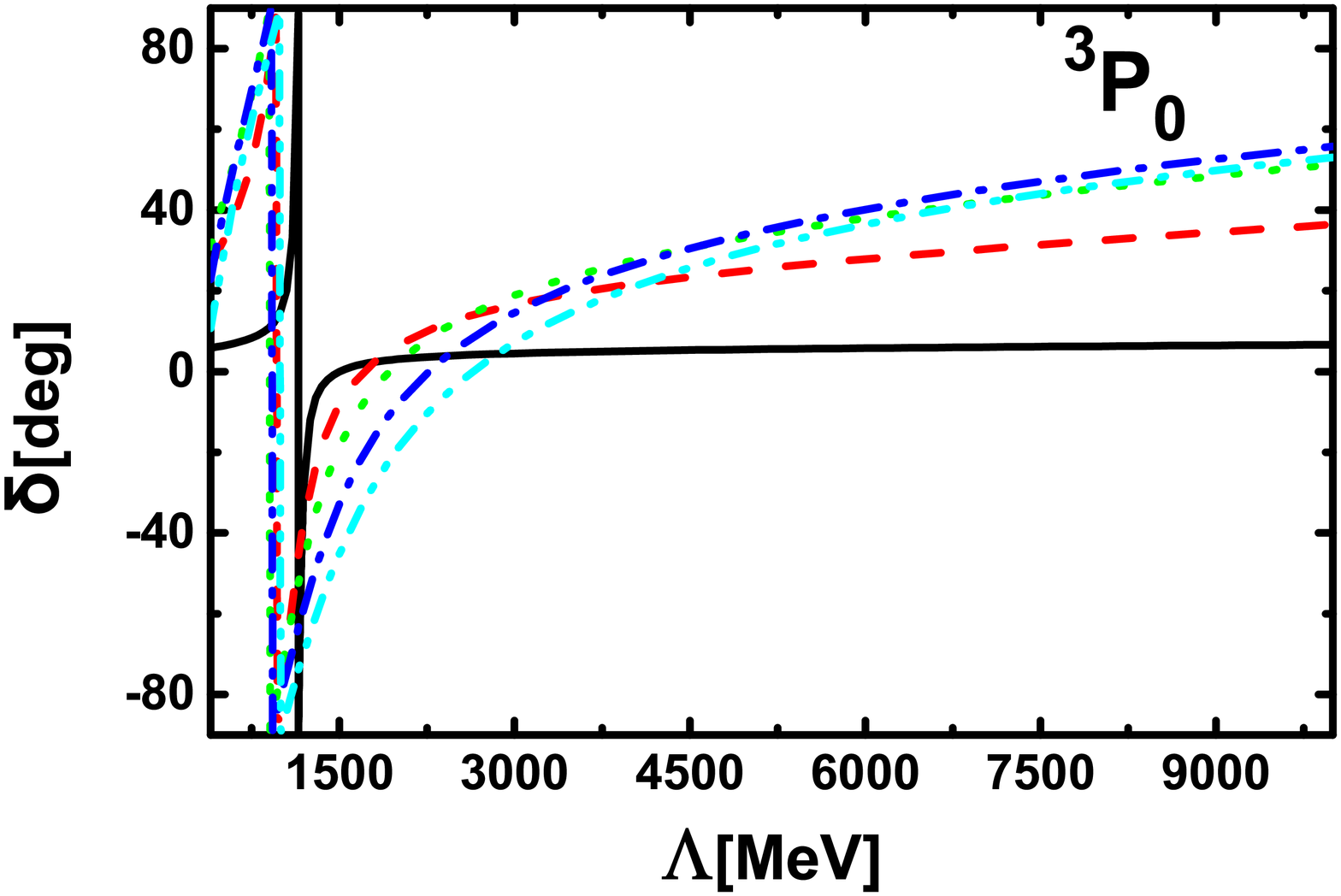}}
&{\includegraphics[width=5.5cm,height=3.5cm]{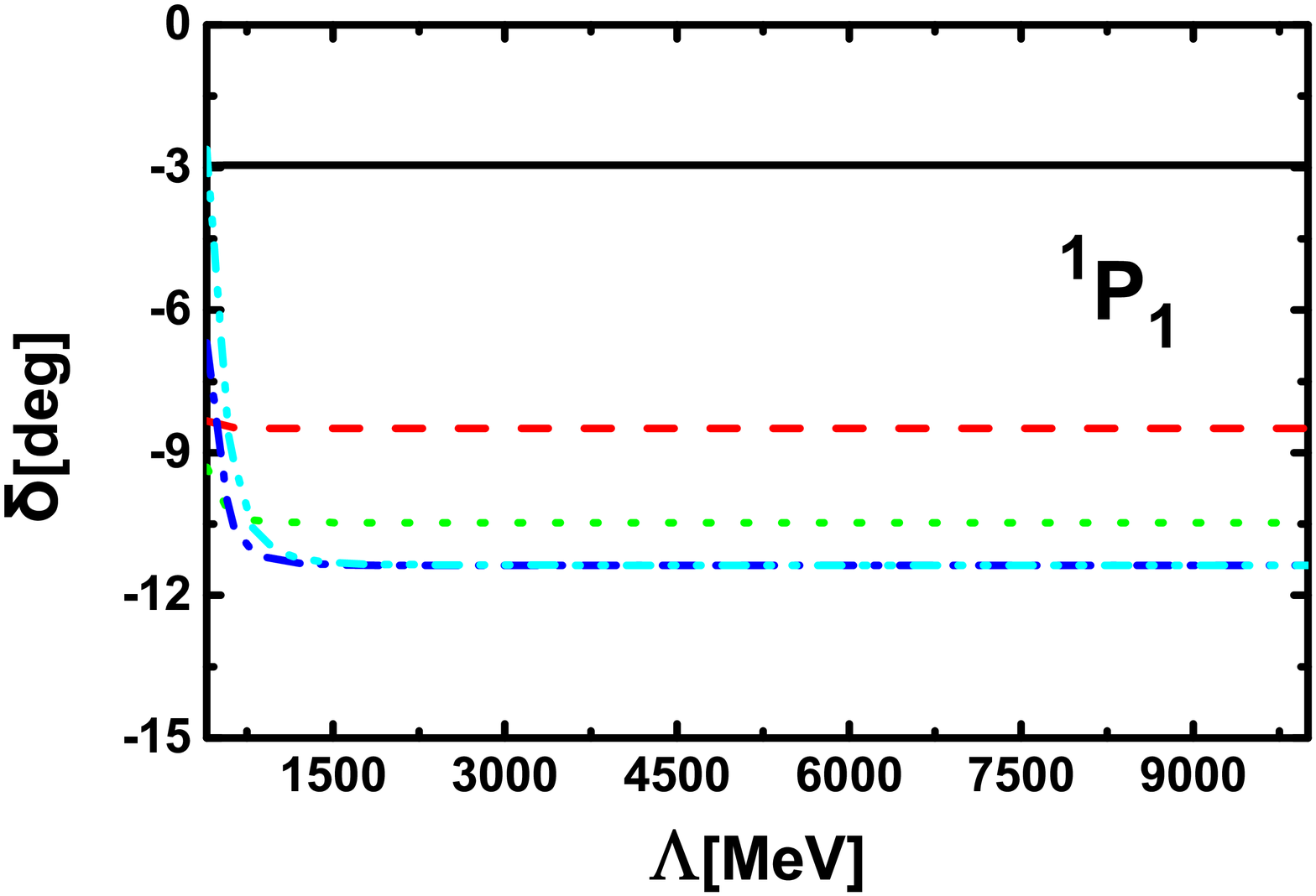}}
&{\includegraphics[width=5.5cm,height=3.5cm]{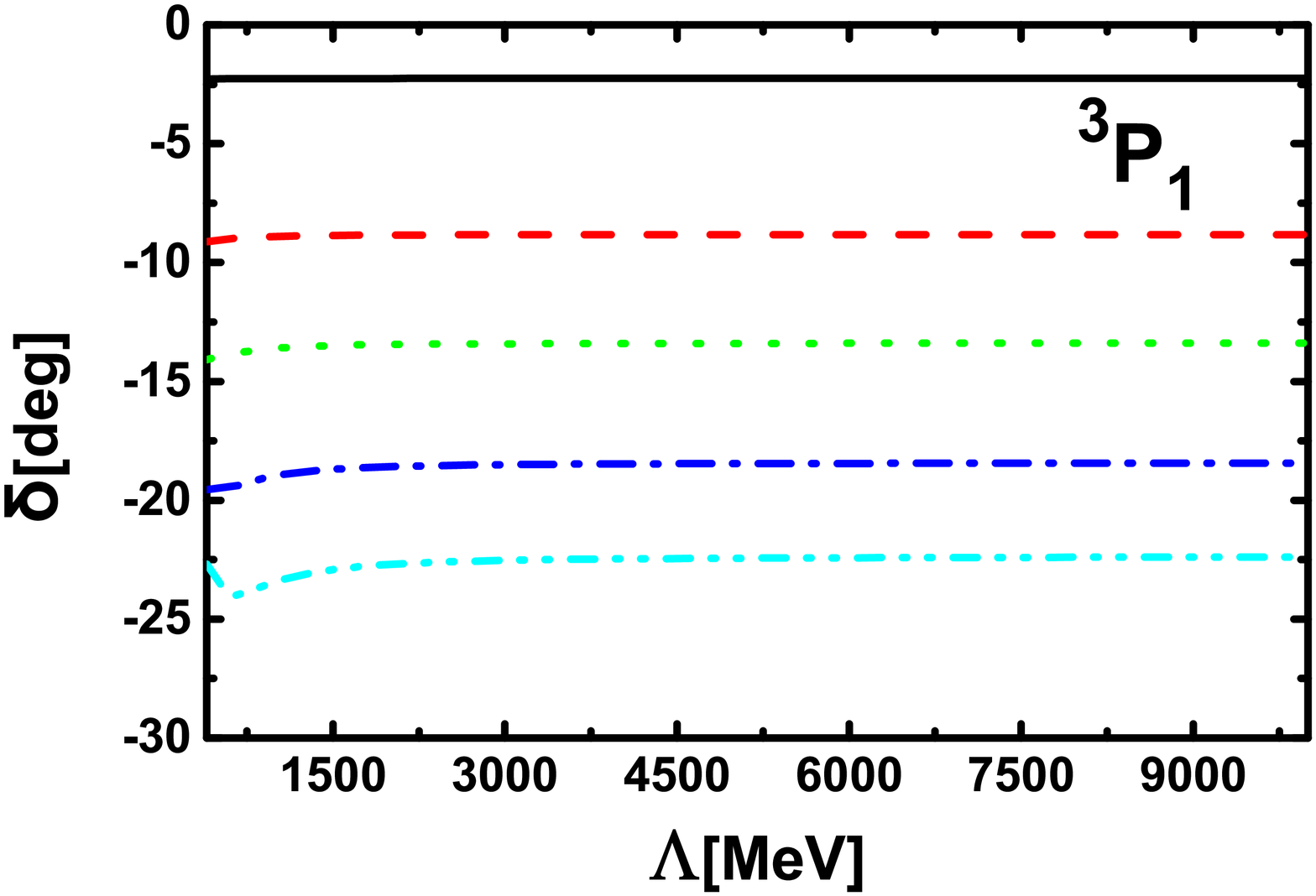}}\\
{\includegraphics[width=5.5cm,height=3.5cm]{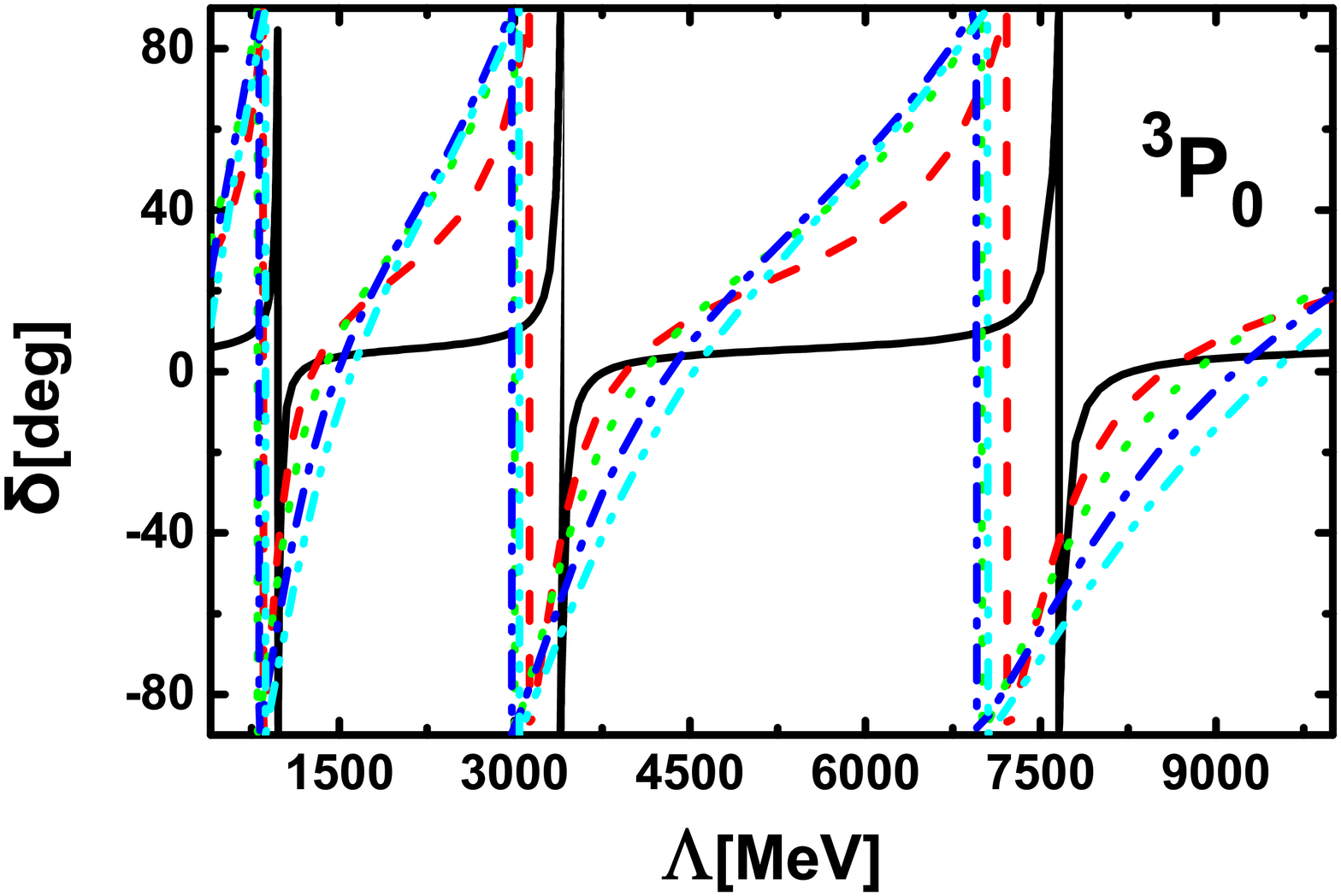}}
&{\includegraphics[width=5.5cm,height=3.5cm]{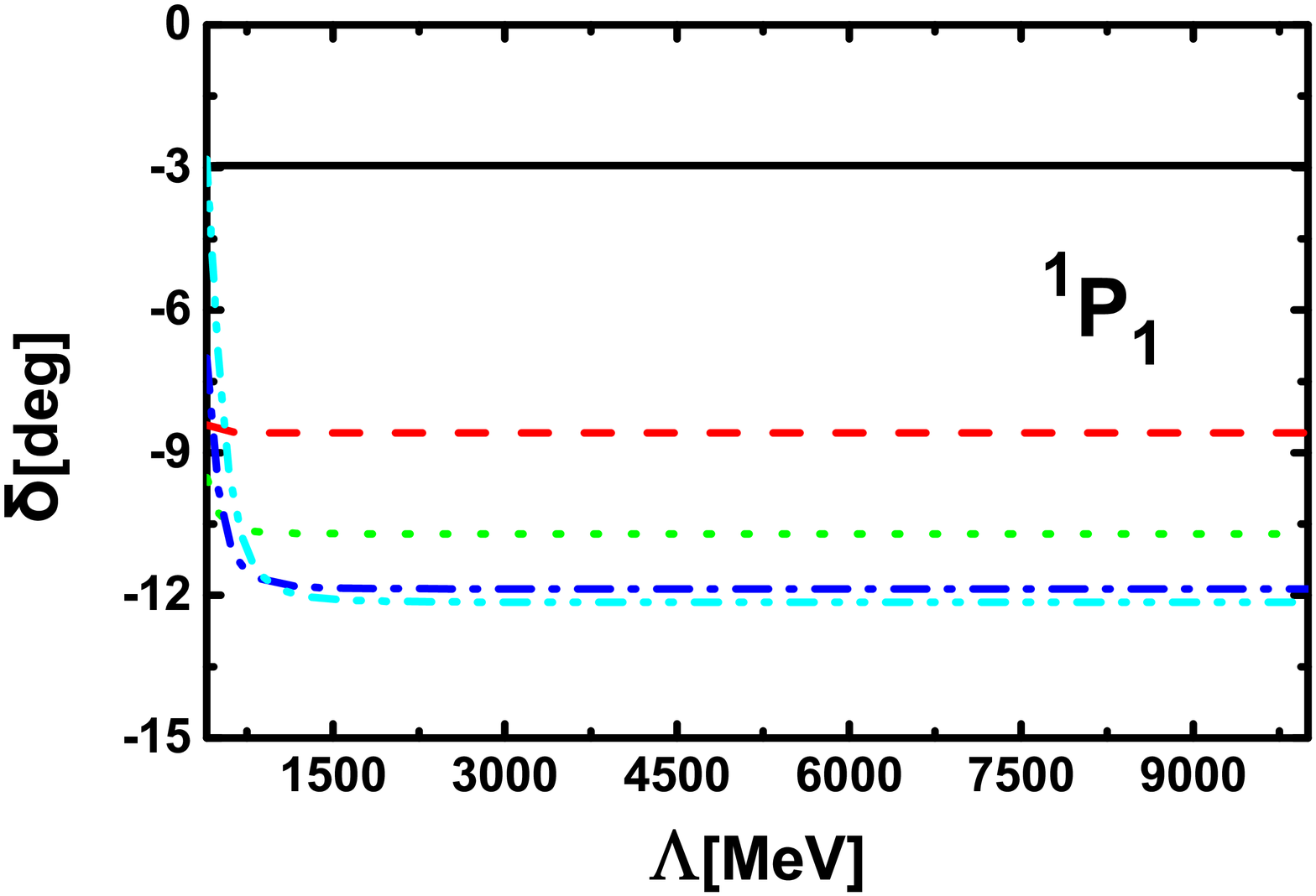}}
&{\includegraphics[width=5.5cm,height=3.5cm]{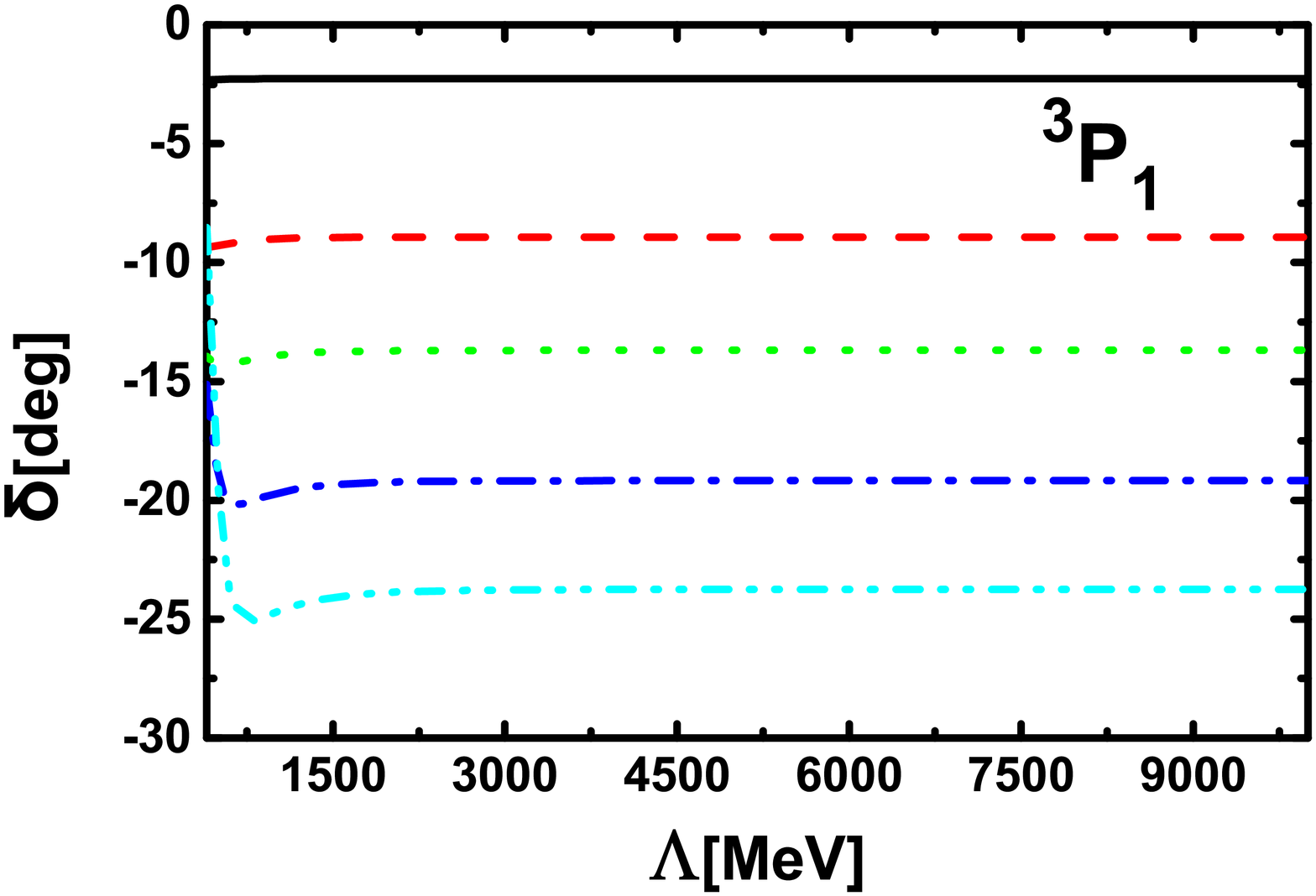}}
\end{tabular}
\caption{Comparison of the $^3P_{0}$, $^1P_{1}$, and $^3P_{1}$ phase shifts (as functions of the $\Lambda$) for laboratory energies of 10 MeV (black solid lines), 50 MeV (red dashed lines), 100 MeV (green dotted lines), 190 MeV (blue dash-dotted lines), 300 MeV (cyan dash-dot-dotted lines). Phase shifts in the upper row are obtained in the covariant scheme, while
 those in the lower row are obtained in the Weinberg scheme. Note that only the OPE contribution is considered. }\label{fig:Compare-OPE}.
\end{figure}

It is instructive to compare at LO the covariant framework with
%
%
the non-relativistic one, on which Weinberg counting is based, when only the long-range force present --- OPE---is present.  The phase shifts for different laboratory energies as a function of the cutoff are shown in  Fig.~\ref{fig:Compare-OPE}. It is clear that the OPE is cutoff independent for the
$^1P_1$ and $^3P_1$ channels, while it is not for the $^3P_0$ channel, where a limit-cycle-like behavior appears in both approaches.
However, as already noticed in Ref.~\cite{Li:2016paq}, the interval between adjacent cycles is bigger in the Kadyshevsky equation (used in
the covariant scheme) than in the
Lippmann-Schwinger approach (used in the Weinberg approach). In the present case, the second cycle appears at $\Lambda=10.6$ GeV in the covariant scheme.

We note by passing that although the OPE in $^3P_1$ is cutoff independent, once a contact term is added and fixed by fitting to the corresponding phase shift, this channel becomes cutoff dependent, both in the present case and in Ref.~\cite{Yang:2009kx}.

\subsection{Nucleon-nucleon phase shifts for $^3P_{0}$}
\begin{figure}
\centering
\begin{tabular}{cc}
{\includegraphics[scale=0.25]{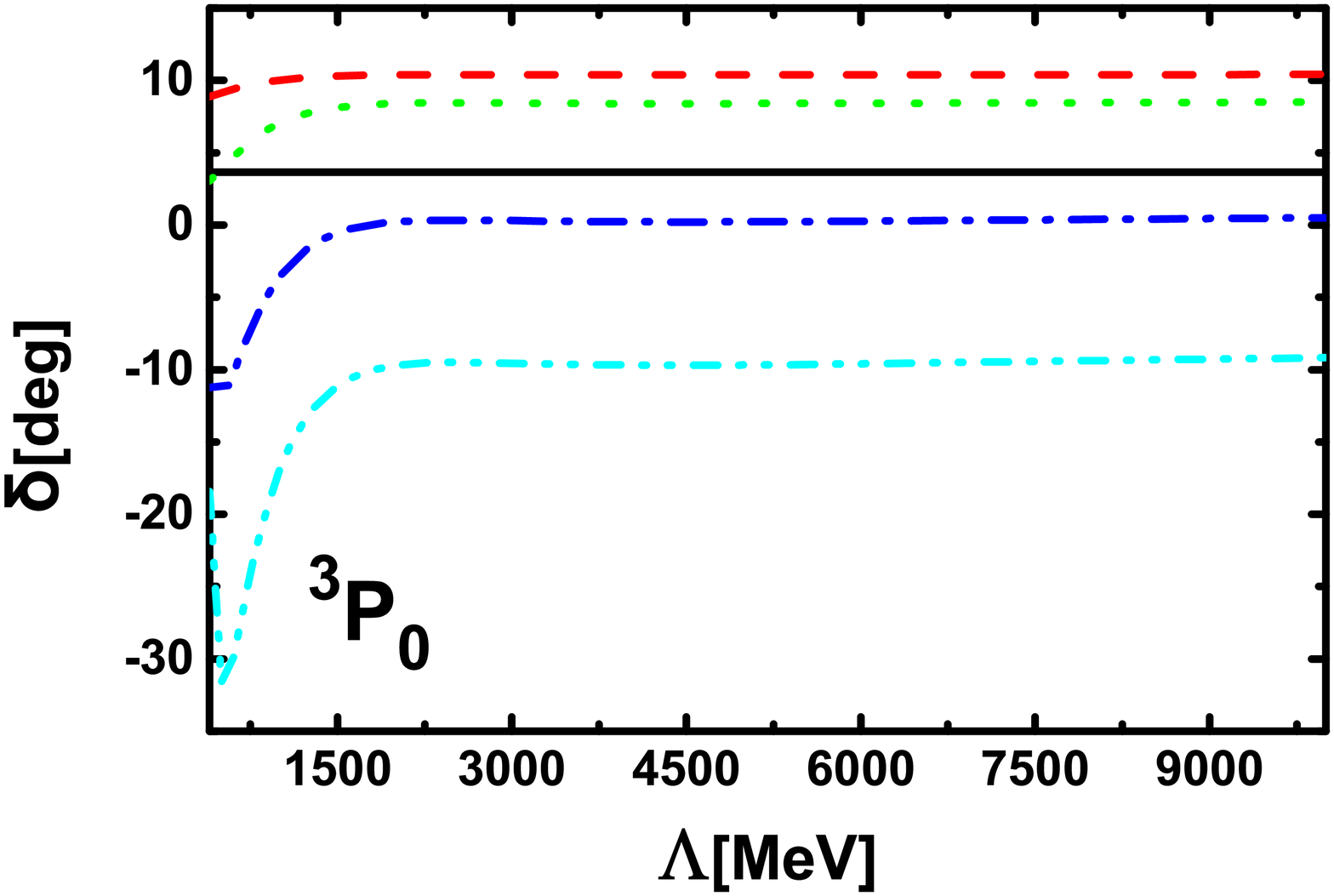}}&{\includegraphics[scale=0.25]{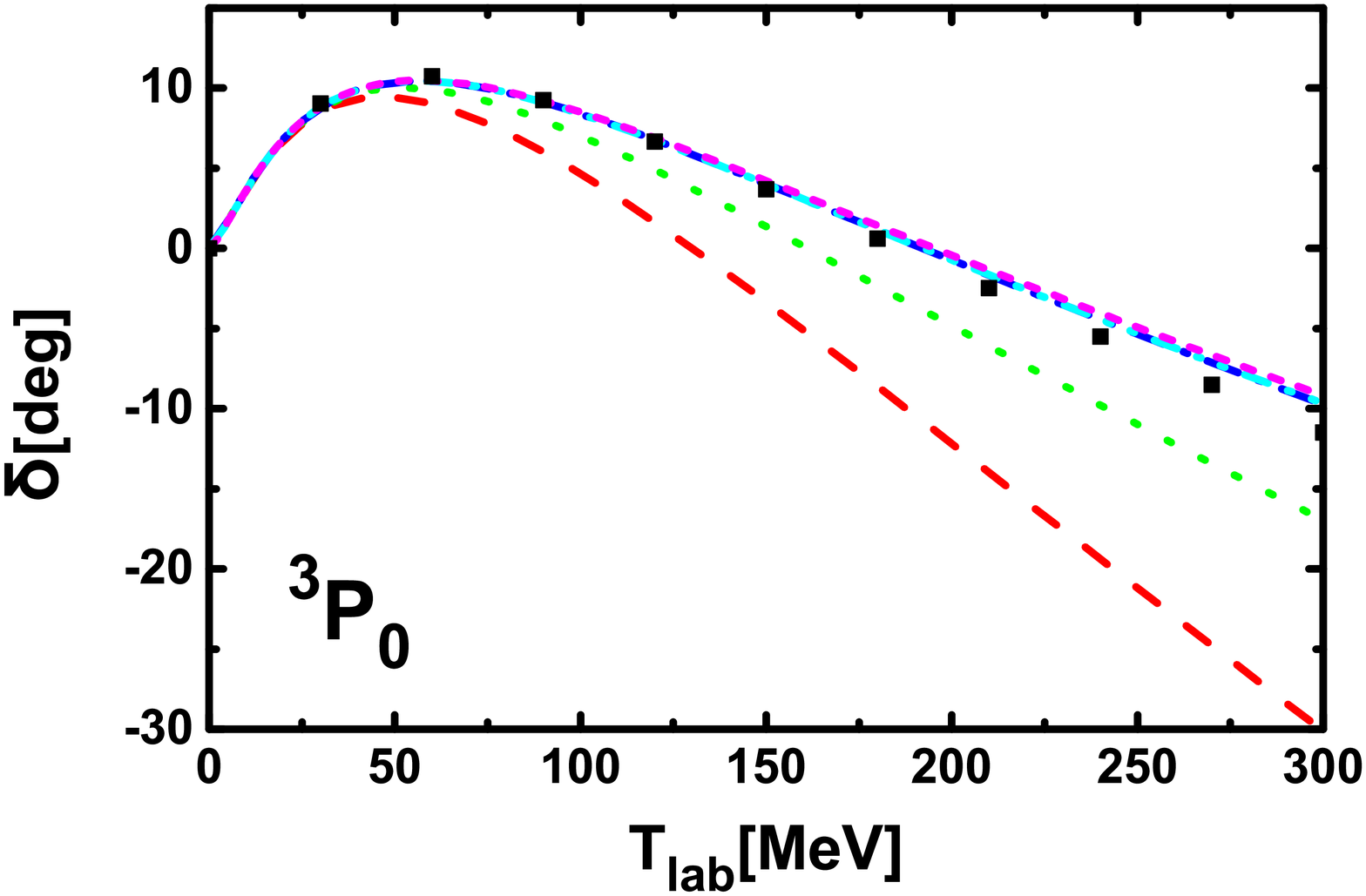}}
\end{tabular}
\caption{Phase shifts as functions of the cutoff $\Lambda$  for laboratory energies of 10 MeV (black solid line), 50 MeV (red dashed line), 100 MeV (green dotted line), 190 MeV (blue dash-dotted line), 300 MeV (cyan dash-dot-dotted line) and  as functions of laboratory energies  with $\Lambda$ of 600  MeV (red dashed line), 1000  MeV (green dotted line), 2000  MeV (blue dash-dotted line), 5000 MeV (cyan dash-dot-dotted line), 10000 MeV (magenta short dashed line). The black diamonds are the Nijmegen phase shifts~\cite{Stoks:1993tb}. }\label{fig:3P0}
\end{figure}

We first discuss the much debate $^3P_0$ channel, where in the Weinberg scheme RGI is lost. In our covariant scheme,
the $^3P_{0}$ channel is not coupled to any other channel and the corresponding contact potential  is given in Eq.(\ref{eq:con}). The
phase shifts as functions of $\Lambda$ and laboratory energies are shown in Fig.~\ref{fig:3P0}. Clearly, the dependence on $\Lambda$ becomes weaker and weaker with increasing $\Lambda$ even for $T_\mathrm{lab.}$ up to 300 MeV. Form
the perspective of RGI, the covariant PC is consistent in this channel. The agreement is good  up to  $T_\mathrm{lab.}=200$ MeV. One should note that in Ref.~\cite{Epelbaum:2012ua}, such a term was promoted to leading order in order to have RGI in this channel.

\subsection{Nucleon-nucleon phase shifts for $^1S_{0}$ and $^3P_{1}$\label{sec-1S03P1}}

In non-relativistic pion-less EFT, the Wigner bound~\cite{Wigner:1955zz} constrains more strongly the value taken by the effective range at higher cutoffs~\cite{Phillips:1996ae,Phillips:1997xu}. We have observed numerically a similar bound in $^1S_{0}$ with the covariant integral equation ~\eqref{eq:Kady}: for cutoff values higher than $\sim 650$ MeV, the $^1S_{0}$ scattering length and effective range can not be fitted simultaneously to their empirical values. Regarding the previously stated fitting strategy, this means that, for high enough cutoff values, we can no longer make predictions on $^3P_{1}$ from $^1S_{0}$ inputs. With $C_{3P1}$ to be fitted to $^3P_{1}$ phase shifts, RGI is lost because OPE is singularly repulsive in this partial wave~\cite{Yang:2009kx}. We note that one solution is suggested by recent works: perturbation theory in all partial waves except for $^1S_0$, $^3S_1$-$^3D_1$, and $^3P_0$~\cite{Wu:2018lai,Kaplan:2019znu}.

\begin{figure}
\centering
\begin{tabular}{cc}
{\includegraphics[scale=0.25]{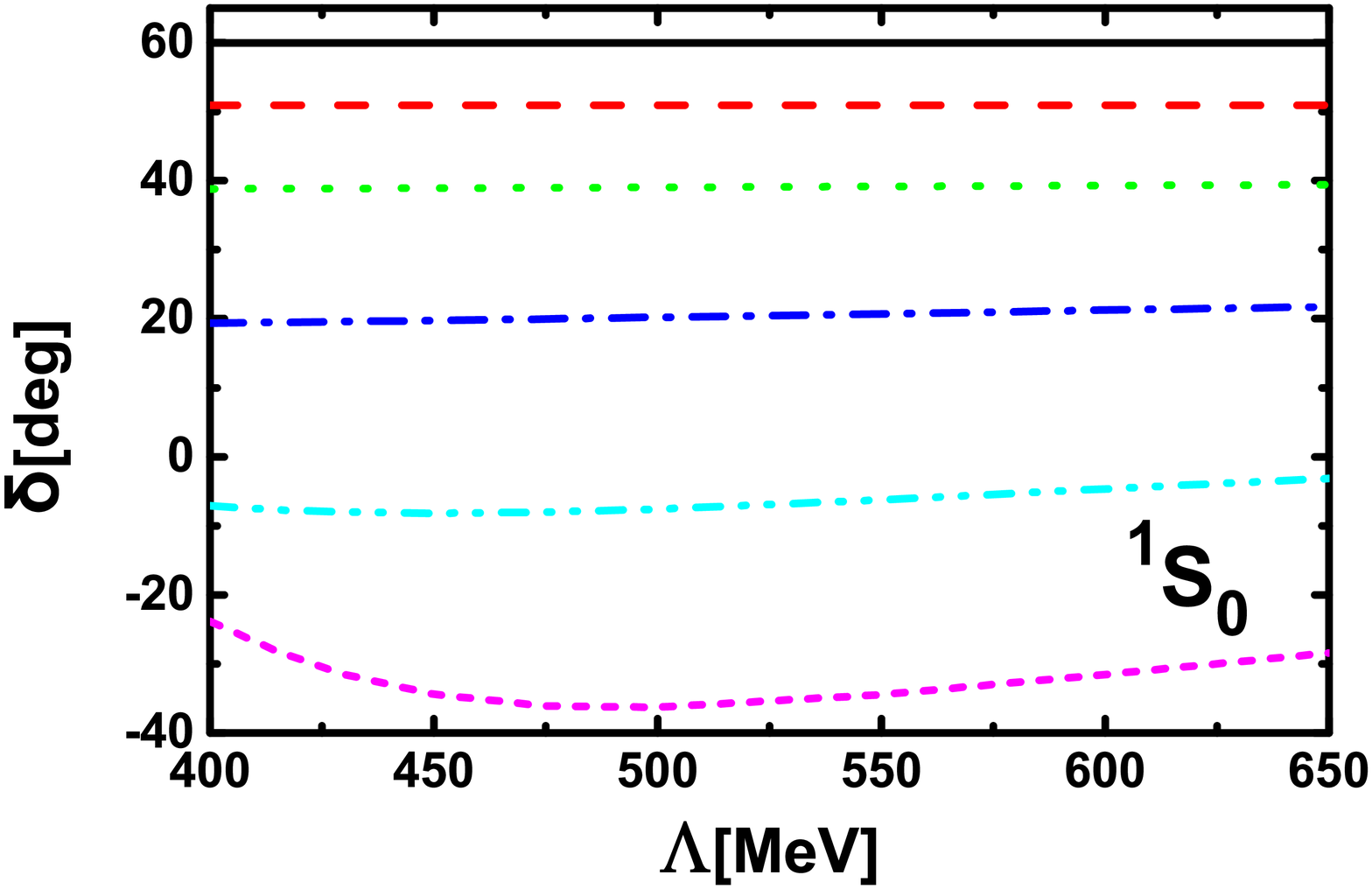}}&{\includegraphics[scale=0.25]{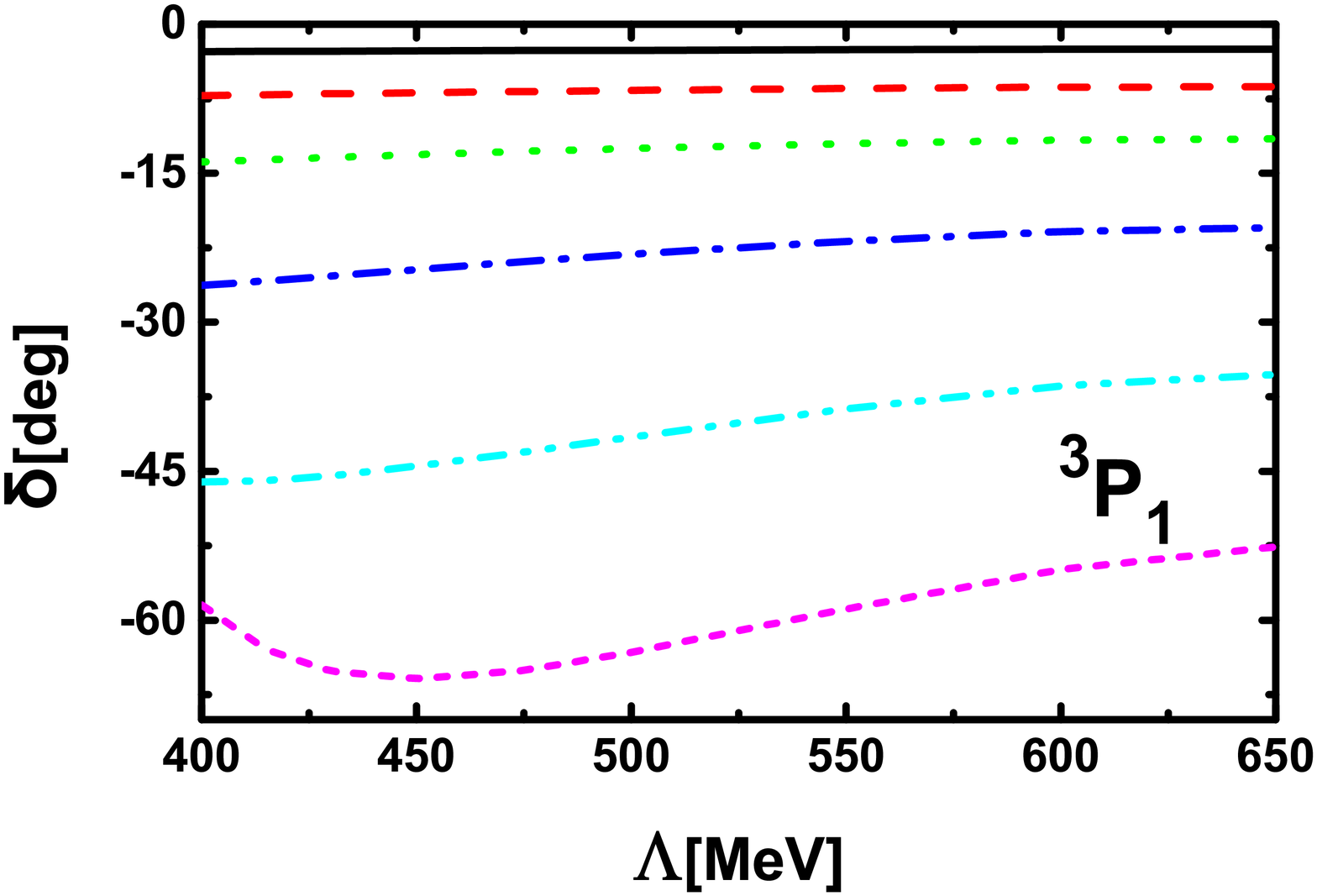}}
\end{tabular}
\caption{Cutoff dependence of the $^1S_0$ and $^3P_1$  phase shifts
for laboratory energies of 10 MeV (black solid lines), 25 MeV (red dashed lines), 50 MeV (green dotted lines), 100 MeV (blue dash-dotted lines), 190 MeV (cyan dash-dot-dotted lines), and 300 MeV (magenta short dashed lines). }\label{fig:1S0-3P1-COUPLE-LAMBDA}
\end{figure}

\begin{figure}
\centering
\begin{tabular}{cc}
{\includegraphics[scale=0.25]{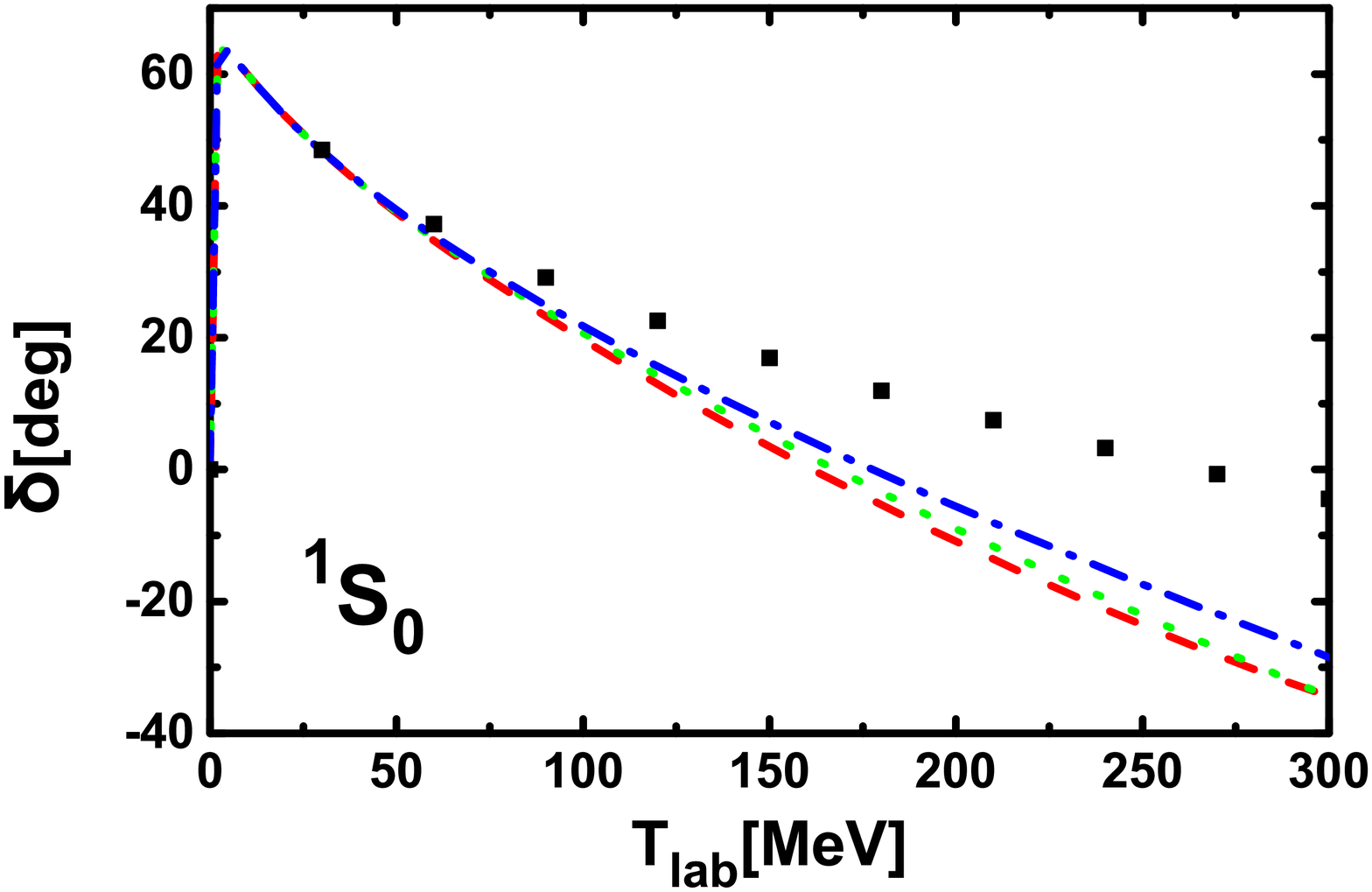}}&{\includegraphics[scale=0.25]{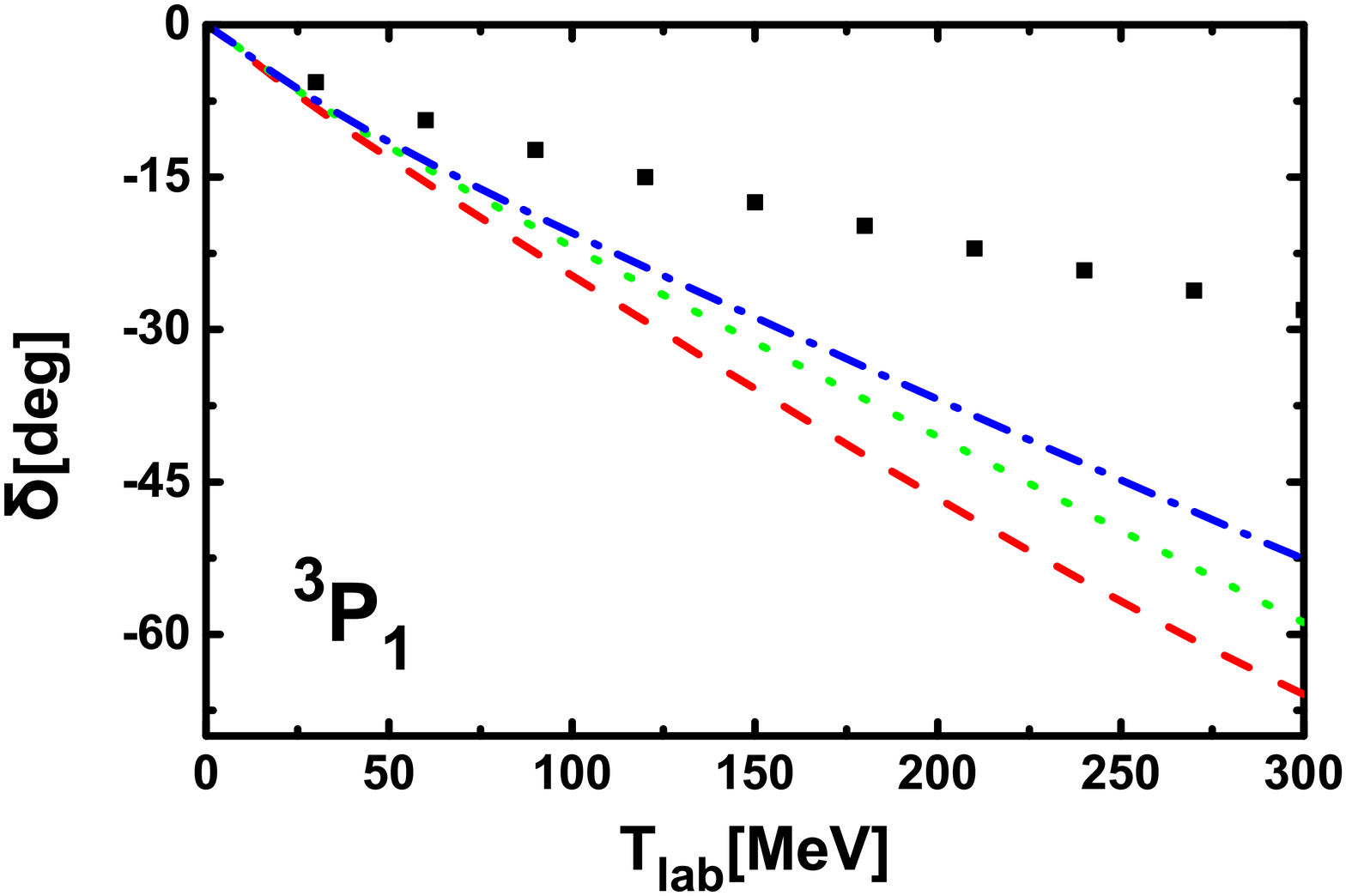}}
\end{tabular}
\caption{Comparison of the $^1S_0$ and $^3P_1$ phase shifts (as functions of the laboratory energy) with $\Lambda$ at 450  MeV (red dashed lines), 550  MeV (green dotted lines), 650  MeV (blue dash-dotted lines) with the Nijmegen phase shifts (black diamonds)~\cite{Stoks:1993tb}.}\label{fig:1S0-3P1-COUPLE-Tlab}
\end{figure}

It is still worth studying these softer cutoffs, so we focus in the following on the region of $\Lambda = 400 - 650$ MeV. From Eq.~(\ref{eq:1s0}), it is clear that the nominally higher order contributions can simulate the
finite nature of the $^1S_0$ potential. With two LECS, we can reproduce the scattering length and effective range simultaneously. This implies that one can describe the corresponding phase shifts better than the LO Weinberg approach, as verified numerically in Ref.~\cite{Ren:2017yvw}. The phase shifts as functions of the cutoff $\Lambda$ for $^1S_{0}$ and $^3P_{1}$ are shown in Fig.~\ref{fig:1S0-3P1-COUPLE-LAMBDA}.  One can see that the dependence on the cutoff in the limited cutoff region is rather weak for both $^1S_0$ and $^3P_1$, Keep in mind that the later is predicted using $C_{3P1}=C_{1S0}-\hat{C}_{1S0}$.
In Fig.~\ref{fig:1S0-3P1-COUPLE-Tlab}, we see that as the cutoff increases from 450 to 650 MeV, the description of the two phase shifts become better.

\subsection{Nucleon-nucleon phase shifts for $^3S_{1}$, $^3D_{1}$, $E_{1}$ and $^1P_{1}$}

\begin{figure}
\centering
\begin{tabular}{cc}
{\includegraphics[scale=0.25]{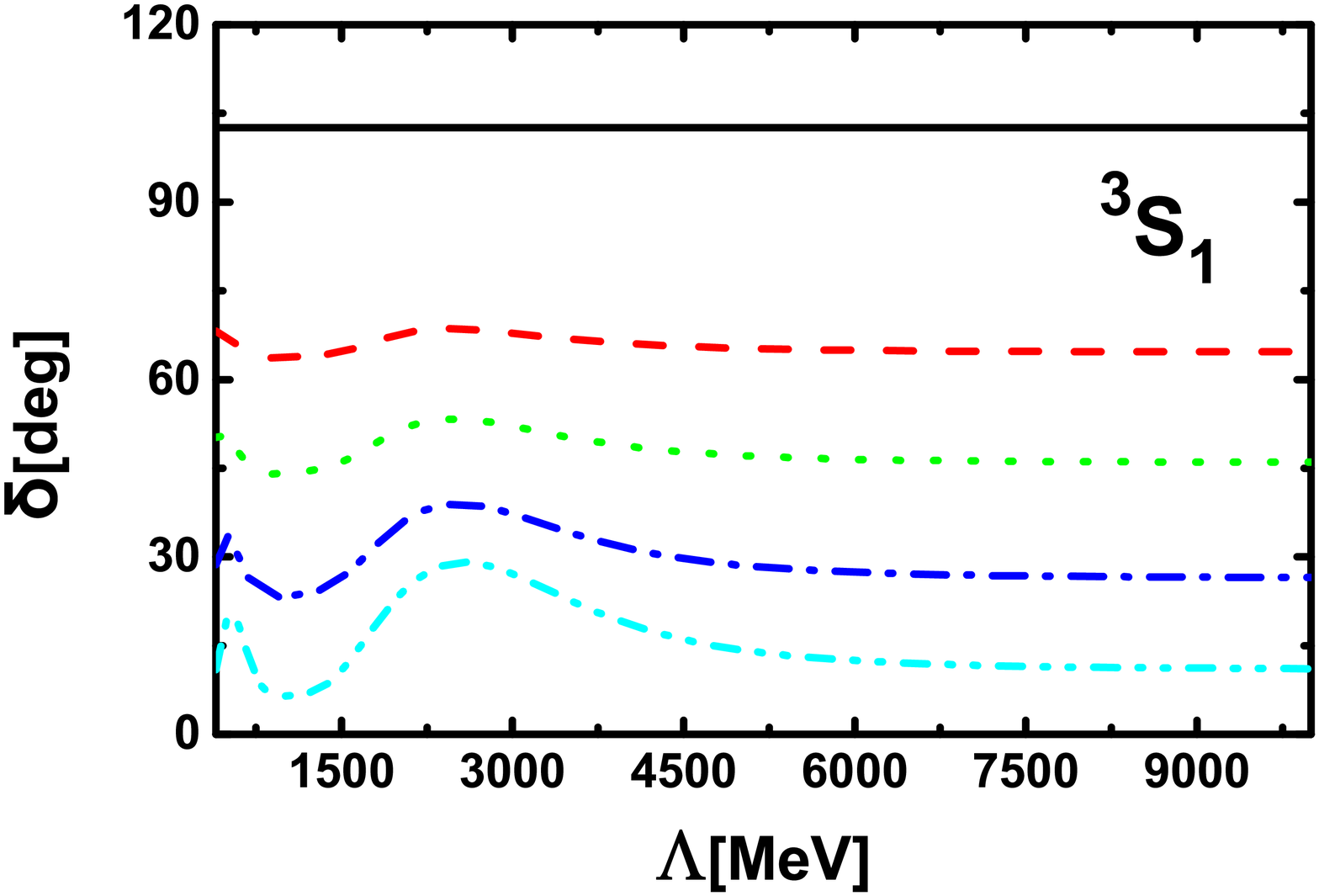}}&{\includegraphics[scale=0.25]{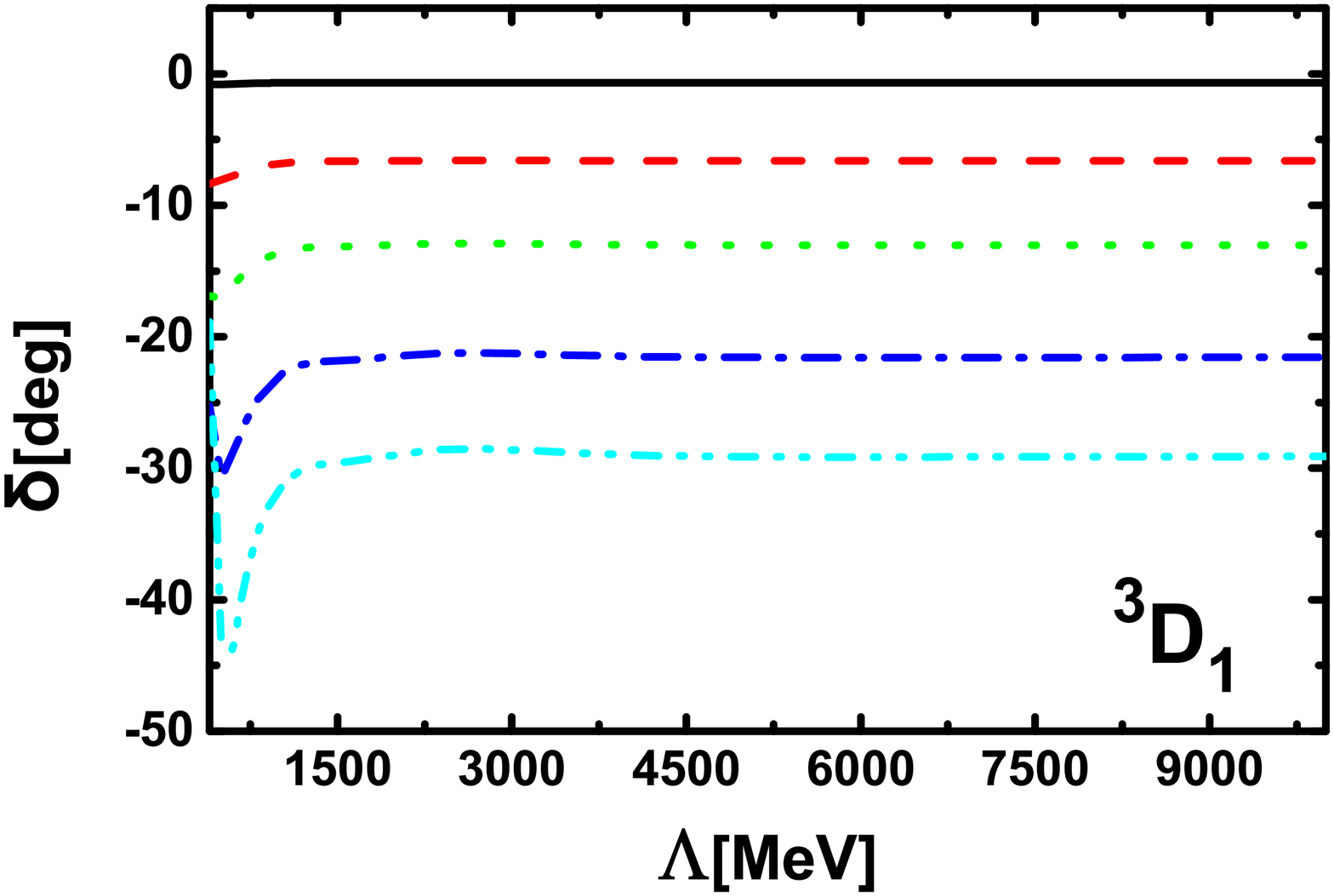}}\\
{\includegraphics[scale=0.25]{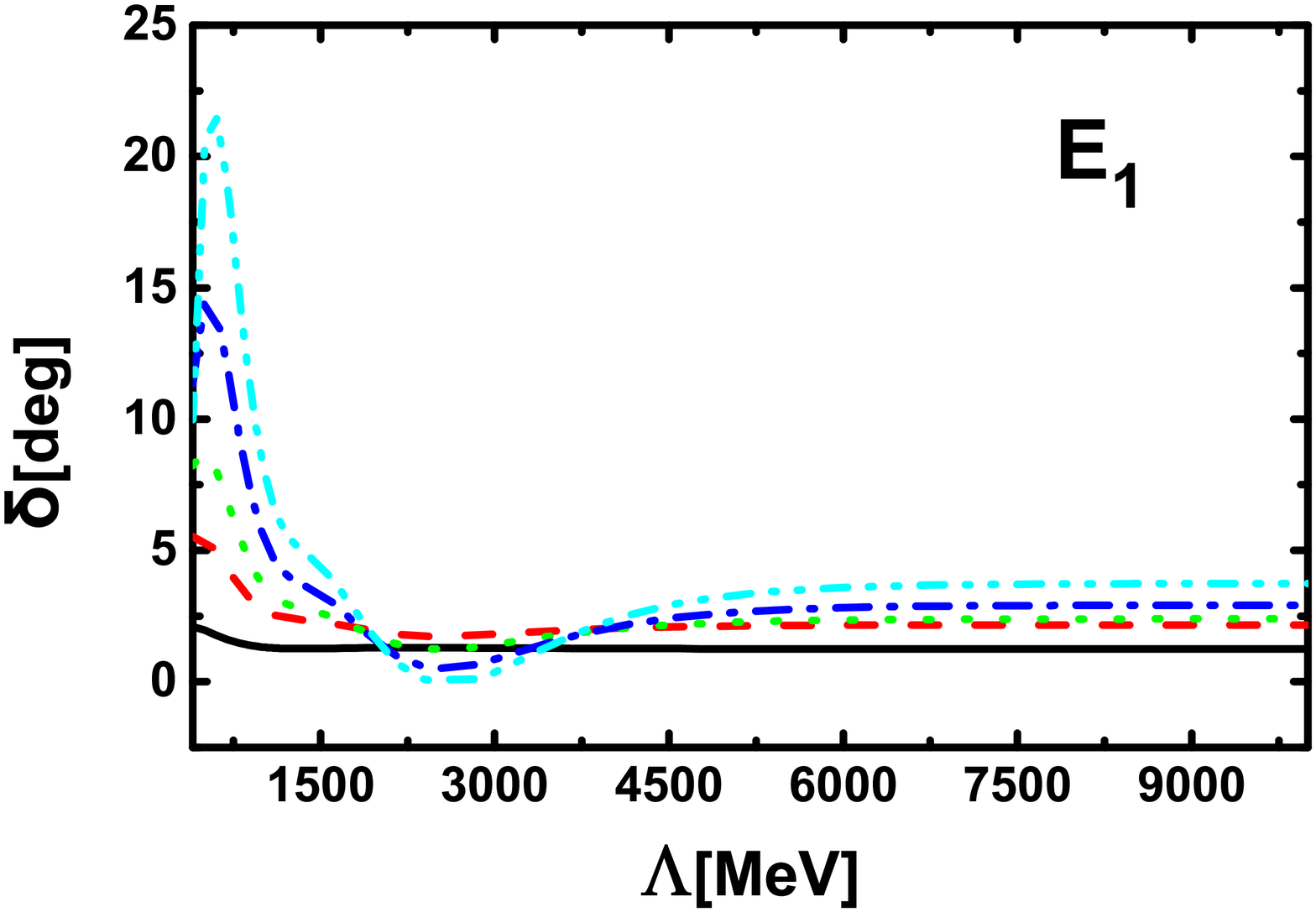}}&{\includegraphics[scale=0.25]{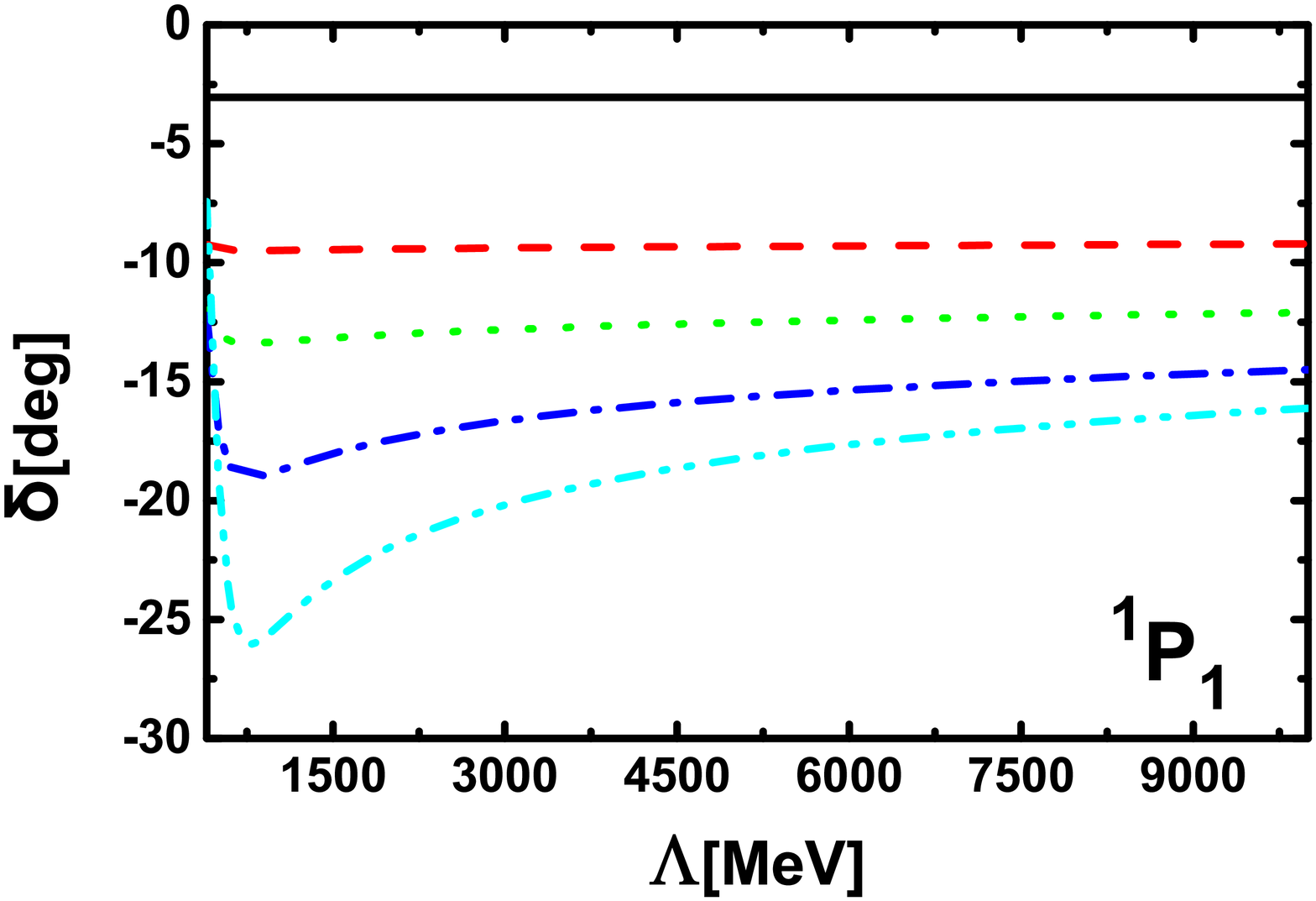}}
\end{tabular}
\caption{The $^3S_{1}$, $^3D_{1}$, $E_{1}$, and $^1P_{1}$ phase shifts for laboratory energies of 10 MeV (black solid lines), 50 MeV (red dashed lines), 100 MeV (green dotted lines), 190 MeV (blue dash-dot line), 300 MeV (cyan dash-dot-dotted lines) as functions of the cutoff $\Lambda$.}\label{fig:3S1-1P1-COUPLE-Lambda}
\end{figure}

The  $^3S_{1}$, $^3D_{1}$, $E_{1}$, and $^1P_{1}$ phase shifts as functions of $\Lambda$ are shown in Fig.~\ref{fig:3S1-1P1-COUPLE-Lambda}. It is seen that the dependence on $\Lambda$ becomes weaker for larger $\Lambda$, indicating
 that in all these channels RGI is satisfied.

\begin{figure}
\centering
\begin{tabular}{cc}
{\includegraphics[scale=0.25]{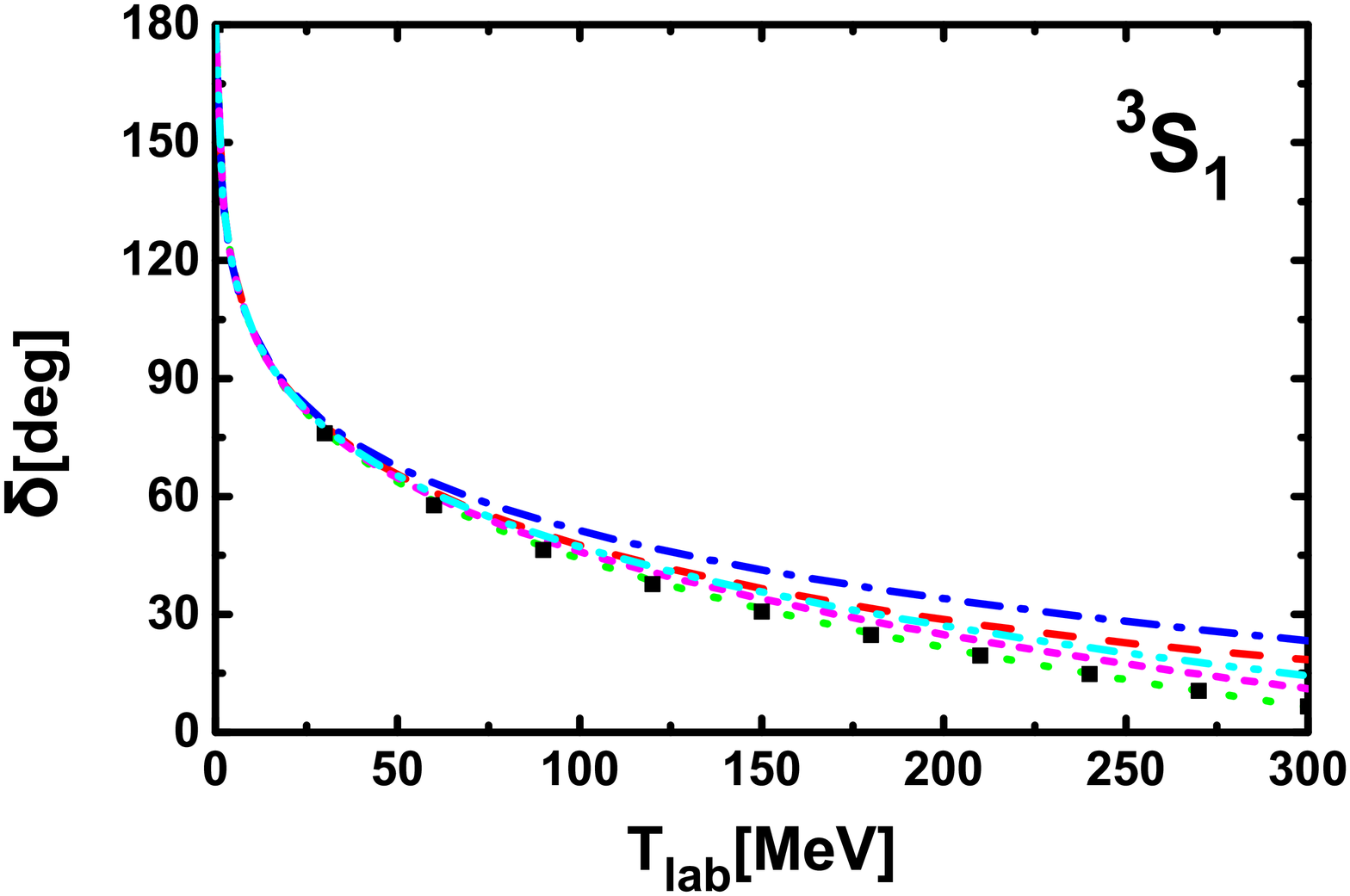}}&{\includegraphics[scale=0.25]{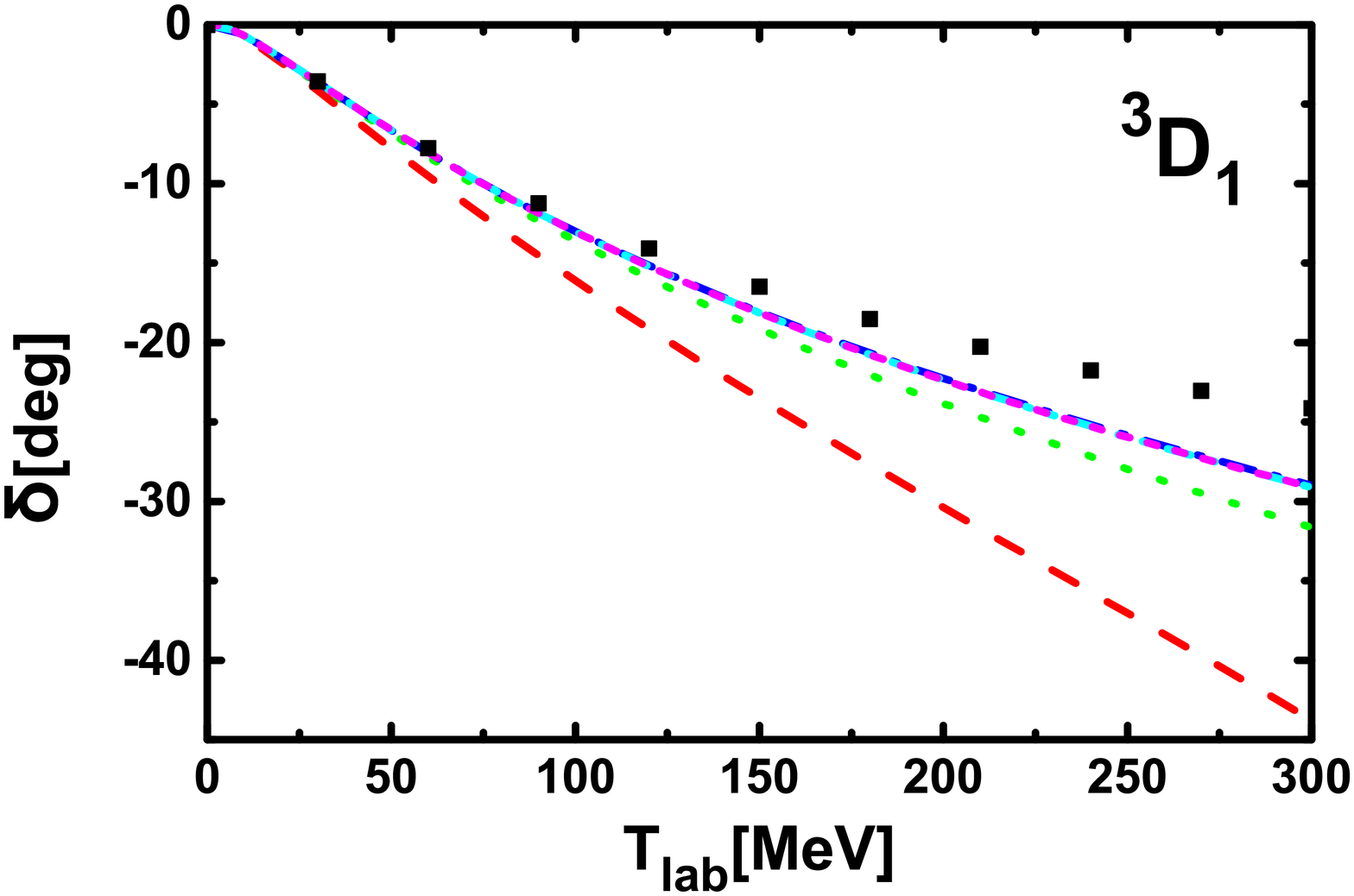}}\\
{\includegraphics[scale=0.25]{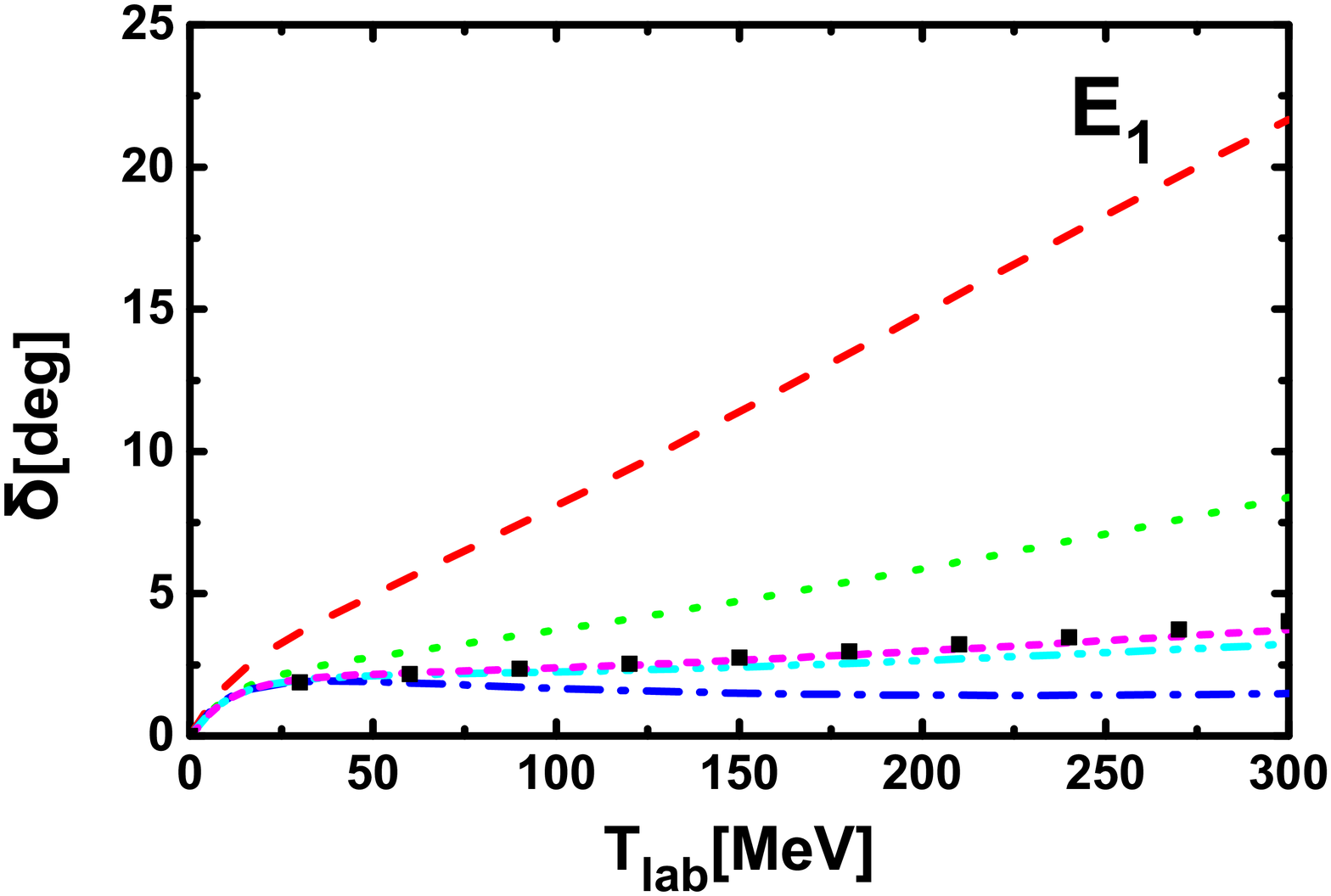}}&{\includegraphics[scale=0.25]{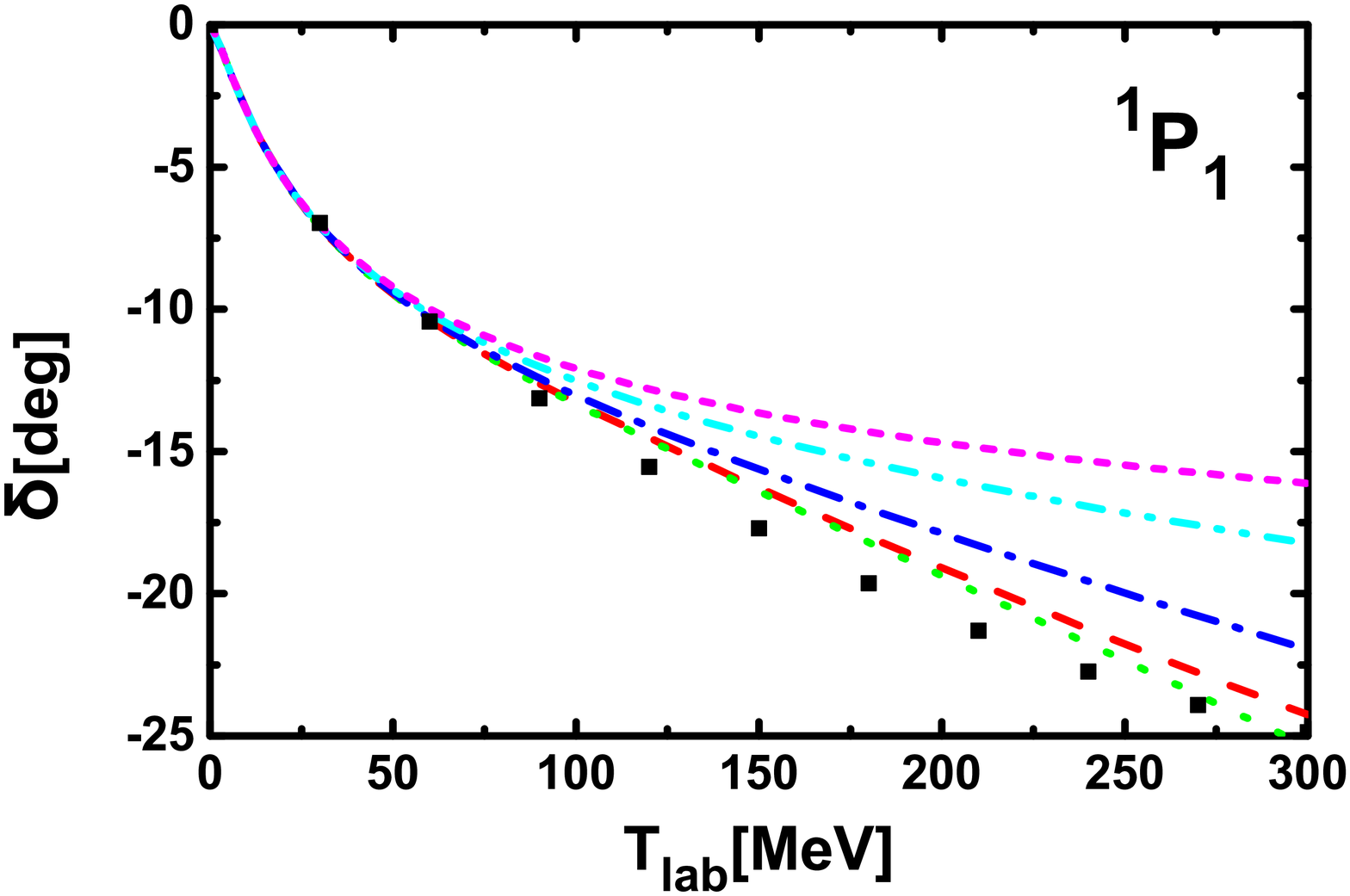}}
\end{tabular}
\caption{Comparison of the $^3S_{1}$, $^3D_{1}$, $E_{1}$, and $^1P_{1}$ phase shifts (as function of the laboratory energy)  with $\Lambda$ at 600  MeV (red dashed lines), 1000  MeV (green dotted lines), 2000  MeV (blue dash-dotted lines), 5000 MeV (cyan dash-dot-dotted lines), 10000 MeV (magenta short dashed lines) with the Nijmegen phase shifts (black diamonds)~\cite{Stoks:1993tb}.}\label{fig:3S1-1P1-COUPLE-Tlab}.
\end{figure}

The phase shifts of these four channels are compared with those of the Nijmegen phase shifts in Fig.~\ref{fig:3S1-1P1-COUPLE-Tlab}. For $^3S_{1}$, the agreement is pretty good when the phase shifts converge. For $^1P_{1}$,   the agreement is good below $T_\mathrm{lab.}=70$ MeV. With the cutoff increasing, the deviation becomes larger at high laboratory energies and therefore higher chiral order contributions are needed.   For $E_{1}$, the  agreement with the Nijmegen analysis is quite good even up to $T_\mathrm{lab.}=300$ MeV when it converges. For $^3D_{1}$, the agreement is good up to $T_\mathrm{lab.}<100\textrm{MeV}$.

\section{Summary and outlook}
In this work, we have analyzed renormalization group invariance of the  leading order covariant chiral nucleon-nucleon force. There are five LECS in all the $J=0,1$ channels. We identified the relations among them and checked the consistency of power counting from
the perspective of renormalization group invariance in the $^3S_{1}$, $^3D_{1}$, $E_{1}$, $^3P_{0}$, and $^1P_{1}$ channels. In the much discussed $^3P_0$ channel,
renormalization group invariance is automatically satisfied in
the covariant power counting as well. On the other hand, the $^1S_{0}$ and $^3P_{1}$ channels are correlated. Therefore, we fix the LECs $C_{1S0}$ and $\hat{C}_{1S0}$ by fitting to the $^1 S_0$ phase shifts and use the relation $\hat{C}_{1S0}=C_{1S0}-C_{3P1}$ to predict $C_{3P1}$. Since the Wigner bound restricts the maximum cutoff allowed in this channel, we only varied the cutoff in a limited region of 400-650 MeV. The resulting phase shifts in the two channels turn out to be  reasonable.   Similar to
the Weinberg power counting, the $^3S_{1}$, $^3D_{1}$, $E_{1}$ and $^1P_{1}$ channels are renormalization group invariant.

It must be noted that after many years of extensive studies, there is
yet no consensus on the meaning of and no universally accepted solutions to the
non-perturbative renormalization of the pion-ful chiral nuclear force.   The present work should only be
viewed a new attempt at tackling this long-standing problem from a  different perspective. The results shown in
the present work indicate that we are still far away from providing a solution and therefore more works are needed, for instance, a detailed study along the same line at higher chiral orders.

\section{Acknowledgements}

CXW thanks Xiu-Lei Ren for useful discussions at the early stage of the present work. We would like to thank
Manuel Pavon Valderrama and Bira van Kolck for stimulating discussions and comments in various occasions.
This work is partly supported by the National Natural Science Foundation of China under Grant Nos.11735003, 11975041, 11775148, and 11961141004.

\section{Appendix}

\begin{figure}[htpb]
\centering
\begin{tabular}{cc}
{\includegraphics[scale=0.25]{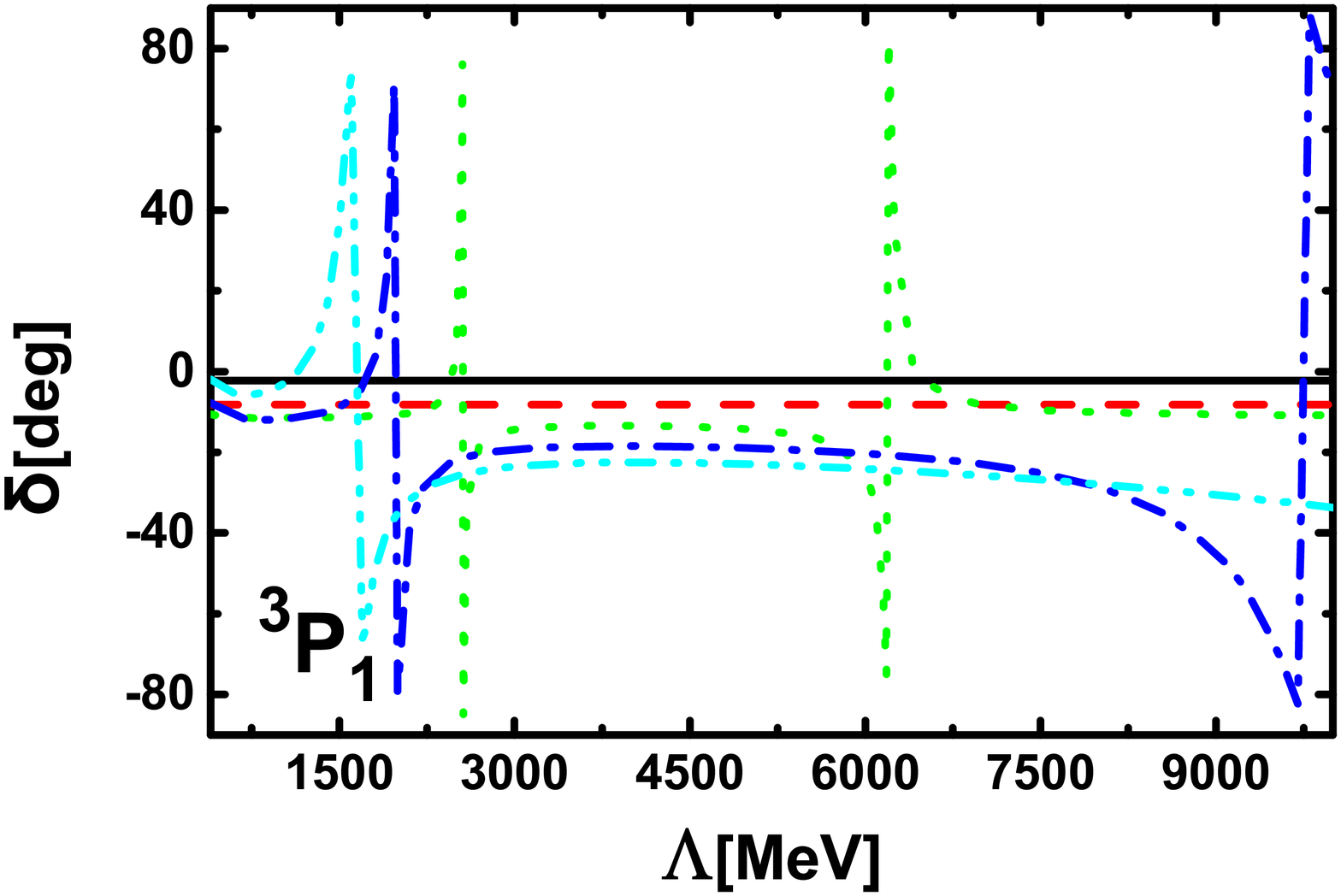}}&{\includegraphics[scale=0.25]{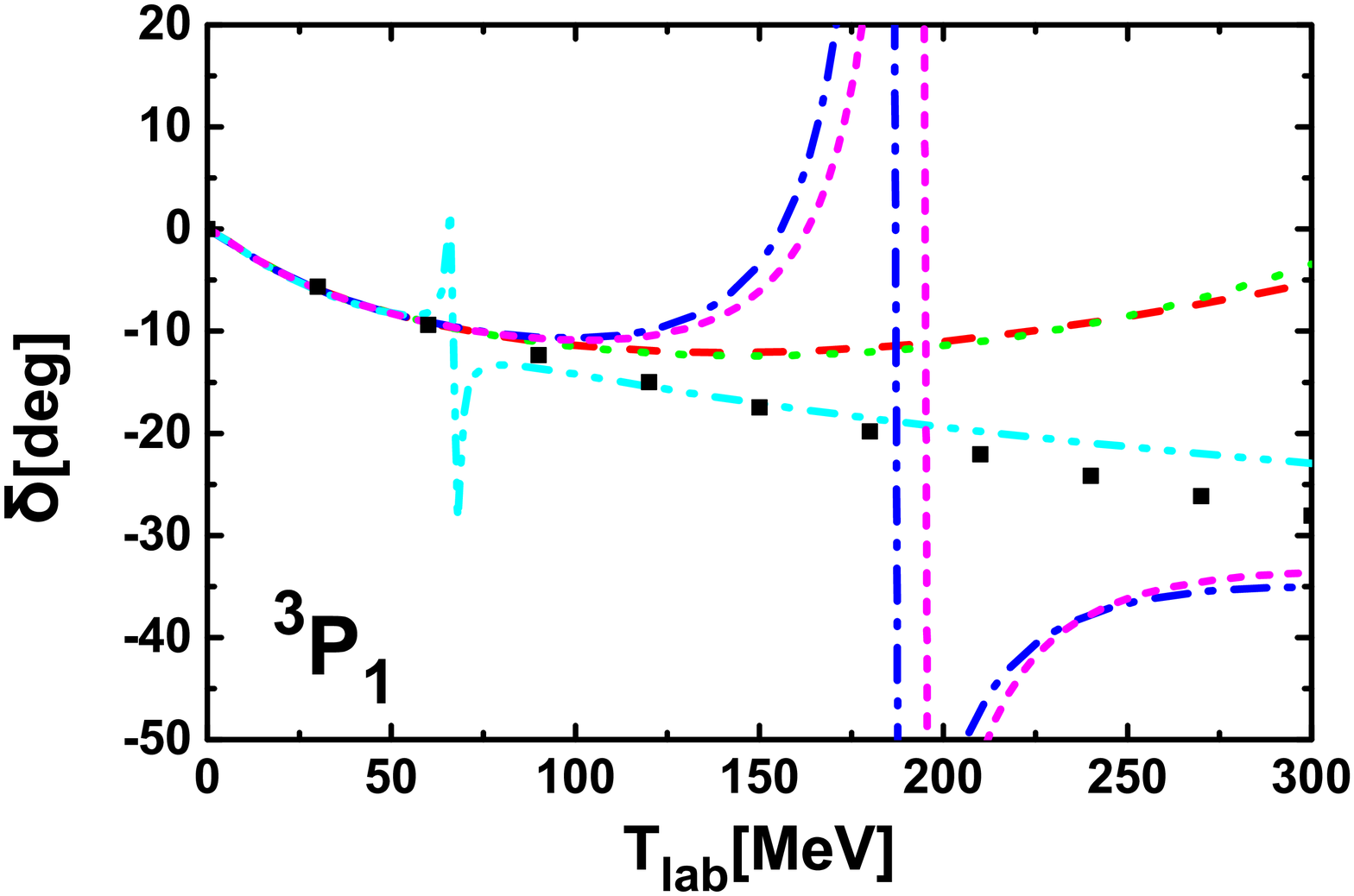}}
\end{tabular}
\caption{Phase shifts as functions of $\Lambda$ for laboratory energy of 10 MeV (black solid line), 50 MeV (red dashed line), 100 MeV (green dotted line), 190 MeV (blue dash-dot line), 300 MeV (cyan dash-dot-dot line) and  as functions of laboratory energies with $\Lambda$ fixed at 600  MeV (red dash line), 1000  MeV (green dot line), 2000  MeV (blue dash-dot line), 5000 MeV (cyan dash-dot-dot line), 10000 MeV (magenta short dashed line). The black diamonds are the Nijmegen phase shifts~\cite{Stoks:1993tb}.}\label{fig:3P1}
\end{figure}

Here, we show the result of $^3P_1$ with $C_{3P1}$ fixed by fitting to the phase shift at 50 MeV in Fig.~\ref{fig:3P1}. It is clear that $^3P_1$ is not renormalization group invariant. This conclusion is similar to Ref.~\cite{Yang:2009pn} where it is shown that contact interactions in $^3P_1$, if treated non-perturbatively, destroy the renormalizability of this channel .

\bibliography{nuclear-force.bib}

\begin{thebibliography}{48}%
\makeatletter
\providecommand \@ifxundefined [1]{%
 \@ifx{#1\undefined}
}%
\providecommand \@ifnum [1]{%
 \ifnum #1\expandafter \@firstoftwo
 \else \expandafter \@secondoftwo
 \fi
}%
\providecommand \@ifx [1]{%
 \ifx #1\expandafter \@firstoftwo
 \else \expandafter \@secondoftwo
 \fi
}%
\providecommand \natexlab [1]{#1}%
\providecommand \enquote  [1]{``#1''}%
\providecommand \bibnamefont  [1]{#1}%
\providecommand \bibfnamefont [1]{#1}%
\providecommand \citenamefont [1]{#1}%
\providecommand \href@noop [0]{\@secondoftwo}%
\providecommand \href [0]{\begingroup \@sanitize@url \@href}%
\providecommand \@href[1]{\@@startlink{#1}\@@href}%
\providecommand \@@href[1]{\endgroup#1\@@endlink}%
\providecommand \@sanitize@url [0]{\catcode `\\12\catcode `\$12\catcode
  `\&12\catcode `\#12\catcode `\^12\catcode `\_12\catcode `\%12\relax}%
\providecommand \@@startlink[1]{}%
\providecommand \@@endlink[0]{}%
\providecommand \url  [0]{\begingroup\@sanitize@url \@url }%
\providecommand \@url [1]{\endgroup\@href {#1}{\urlprefix }}%
\providecommand \urlprefix  [0]{URL }%
\providecommand \Eprint [0]{\href }%
\providecommand \doibase [0]{http://dx.doi.org/}%
\providecommand \selectlanguage [0]{\@gobble}%
\providecommand \bibinfo  [0]{\@secondoftwo}%
\providecommand \bibfield  [0]{\@secondoftwo}%
\providecommand \translation [1]{[#1]}%
\providecommand \BibitemOpen [0]{}%
\providecommand \bibitemStop [0]{}%
\providecommand \bibitemNoStop [0]{.\EOS\space}%
\providecommand \EOS [0]{\spacefactor3000\relax}%
\providecommand \BibitemShut  [1]{\csname bibitem#1\endcsname}%
\let\auto@bib@innerbib\@empty
\bibitem [{\citenamefont {Weinberg}(1990)}]{Weinberg:1990rz}%
  \BibitemOpen
  \bibfield  {author} {\bibinfo {author} {\bibfnamefont {S.}~\bibnamefont
  {Weinberg}},\ }\href {\doibase 10.1016/0370-2693(90)90938-3} {\bibfield
  {journal} {\bibinfo  {journal} {Phys. Lett.}\ }\textbf {\bibinfo {volume}
  {B251}},\ \bibinfo {pages} {288} (\bibinfo {year} {1990})}\BibitemShut
  {NoStop}%
\bibitem [{\citenamefont {Weinberg}(1991)}]{Weinberg:1991um}%
  \BibitemOpen
  \bibfield  {author} {\bibinfo {author} {\bibfnamefont {S.}~\bibnamefont
  {Weinberg}},\ }\href {\doibase 10.1016/0550-3213(91)90231-L} {\bibfield
  {journal} {\bibinfo  {journal} {Nucl. Phys.}\ }\textbf {\bibinfo {volume}
  {B363}},\ \bibinfo {pages} {3} (\bibinfo {year} {1991})}\BibitemShut
  {NoStop}%
\bibitem [{\citenamefont {Epelbaum}\ \emph {et~al.}(2009)\citenamefont
  {Epelbaum}, \citenamefont {Hammer},\ and\ \citenamefont
  {Meissner}}]{Epelbaum:2008ga}%
  \BibitemOpen
  \bibfield  {author} {\bibinfo {author} {\bibfnamefont {E.}~\bibnamefont
  {Epelbaum}}, \bibinfo {author} {\bibfnamefont {H.-W.}\ \bibnamefont
  {Hammer}}, \ and\ \bibinfo {author} {\bibfnamefont {U.-G.}\ \bibnamefont
  {Meissner}},\ }\href {\doibase 10.1103/RevModPhys.81.1773} {\bibfield
  {journal} {\bibinfo  {journal} {Rev. Mod. Phys.}\ }\textbf {\bibinfo {volume}
  {81}},\ \bibinfo {pages} {1773} (\bibinfo {year} {2009})},\ \Eprint
  {http://arxiv.org/abs/0811.1338} {arXiv:0811.1338 [nucl-th]} \BibitemShut
  {NoStop}%
\bibitem [{\citenamefont {Machleidt}\ and\ \citenamefont
  {Entem}(2011)}]{Machleidt:2011zz}%
  \BibitemOpen
  \bibfield  {author} {\bibinfo {author} {\bibfnamefont {R.}~\bibnamefont
  {Machleidt}}\ and\ \bibinfo {author} {\bibfnamefont {D.~R.}\ \bibnamefont
  {Entem}},\ }\href {\doibase 10.1016/j.physrep.2011.02.001} {\bibfield
  {journal} {\bibinfo  {journal} {Phys. Rept.}\ }\textbf {\bibinfo {volume}
  {503}},\ \bibinfo {pages} {1} (\bibinfo {year} {2011})},\ \Eprint
  {http://arxiv.org/abs/1105.2919} {arXiv:1105.2919 [nucl-th]} \BibitemShut
  {NoStop}%
\bibitem [{\citenamefont {Hammer}\ \emph {et~al.}()\citenamefont {Hammer},
  \citenamefont {König},\ and\ \citenamefont {van Kolck}}]{Hammer:2019poc}%
  \BibitemOpen
  \bibfield  {author} {\bibinfo {author} {\bibfnamefont {H.~W.}\ \bibnamefont
  {Hammer}}, \bibinfo {author} {\bibfnamefont {S.}~\bibnamefont {König}}, \
  and\ \bibinfo {author} {\bibfnamefont {U.}~\bibnamefont {van Kolck}},\
  }\href@noop {} {\ }\Eprint {http://arxiv.org/abs/1906.12122}
  {arXiv:1906.12122 [nucl-th]} \BibitemShut {NoStop}%
\bibitem [{\citenamefont {Kaplan}\ \emph {et~al.}(1998)\citenamefont {Kaplan},
  \citenamefont {Savage},\ and\ \citenamefont {Wise}}]{Kaplan:1998tg}%
  \BibitemOpen
  \bibfield  {author} {\bibinfo {author} {\bibfnamefont {D.~B.}\ \bibnamefont
  {Kaplan}}, \bibinfo {author} {\bibfnamefont {M.~J.}\ \bibnamefont {Savage}},
  \ and\ \bibinfo {author} {\bibfnamefont {M.~B.}\ \bibnamefont {Wise}},\
  }\href {\doibase 10.1016/S0370-2693(98)00210-X} {\bibfield  {journal}
  {\bibinfo  {journal} {Phys. Lett.}\ }\textbf {\bibinfo {volume} {B424}},\
  \bibinfo {pages} {390} (\bibinfo {year} {1998})},\ \Eprint
  {http://arxiv.org/abs/nucl-th/9801034} {arXiv:nucl-th/9801034 [nucl-th]}
  \BibitemShut {NoStop}%
\bibitem [{\citenamefont {Yang}()}]{Yang:2019hkn}%
  \BibitemOpen
  \bibfield  {author} {\bibinfo {author} {\bibfnamefont {C.~J.}\ \bibnamefont
  {Yang}},\ }\href@noop {} {\ }\Eprint {http://arxiv.org/abs/1905.12510}
  {arXiv:1905.12510 [nucl-th]} \BibitemShut {NoStop}%
\bibitem [{\citenamefont {Weinberg}(1979)}]{Weinberg:1978kz}%
  \BibitemOpen
  \bibfield  {author} {\bibinfo {author} {\bibfnamefont {S.}~\bibnamefont
  {Weinberg}},\ }\href {\doibase 10.1016/0378-4371(79)90223-1} {\bibfield
  {journal} {\bibinfo  {journal} {Physica}\ }\textbf {\bibinfo {volume}
  {A96}},\ \bibinfo {pages} {327} (\bibinfo {year} {1979})}\BibitemShut
  {NoStop}%
\bibitem [{\citenamefont {Gasser}\ and\ \citenamefont
  {Leutwyler}(1984)}]{Gasser:1983yg}%
  \BibitemOpen
  \bibfield  {author} {\bibinfo {author} {\bibfnamefont {J.}~\bibnamefont
  {Gasser}}\ and\ \bibinfo {author} {\bibfnamefont {H.}~\bibnamefont
  {Leutwyler}},\ }\href {\doibase 10.1016/0003-4916(84)90242-2} {\bibfield
  {journal} {\bibinfo  {journal} {Annals Phys.}\ }\textbf {\bibinfo {volume}
  {158}},\ \bibinfo {pages} {142} (\bibinfo {year} {1984})}\BibitemShut
  {NoStop}%
\bibitem [{\citenamefont {Ecker}\ \emph {et~al.}(1989)\citenamefont {Ecker},
  \citenamefont {Gasser}, \citenamefont {Leutwyler}, \citenamefont {Pich},\
  and\ \citenamefont {de~Rafael}}]{Ecker:1989yg}%
  \BibitemOpen
  \bibfield  {author} {\bibinfo {author} {\bibfnamefont {G.}~\bibnamefont
  {Ecker}}, \bibinfo {author} {\bibfnamefont {J.}~\bibnamefont {Gasser}},
  \bibinfo {author} {\bibfnamefont {H.}~\bibnamefont {Leutwyler}}, \bibinfo
  {author} {\bibfnamefont {A.}~\bibnamefont {Pich}}, \ and\ \bibinfo {author}
  {\bibfnamefont {E.}~\bibnamefont {de~Rafael}},\ }\href {\doibase
  10.1016/0370-2693(89)91627-4} {\bibfield  {journal} {\bibinfo  {journal}
  {Phys. Lett.}\ }\textbf {\bibinfo {volume} {B223}},\ \bibinfo {pages} {425}
  (\bibinfo {year} {1989})}\BibitemShut {NoStop}%
\bibitem [{\citenamefont {Ecker}(1995)}]{Ecker:1994gg}%
  \BibitemOpen
  \bibfield  {author} {\bibinfo {author} {\bibfnamefont {G.}~\bibnamefont
  {Ecker}},\ }\href {\doibase 10.1016/0146-6410(95)00041-G} {\bibfield
  {journal} {\bibinfo  {journal} {Prog. Part. Nucl. Phys.}\ }\textbf {\bibinfo
  {volume} {35}},\ \bibinfo {pages} {1} (\bibinfo {year} {1995})},\ \Eprint
  {http://arxiv.org/abs/hep-ph/9501357} {arXiv:hep-ph/9501357 [hep-ph]}
  \BibitemShut {NoStop}%
\bibitem [{\citenamefont {Pich}(1995)}]{Pich:1995bw}%
  \BibitemOpen
  \bibfield  {author} {\bibinfo {author} {\bibfnamefont {A.}~\bibnamefont
  {Pich}},\ }\href {\doibase 10.1088/0034-4885/58/6/001} {\bibfield  {journal}
  {\bibinfo  {journal} {Rept. Prog. Phys.}\ }\textbf {\bibinfo {volume} {58}},\
  \bibinfo {pages} {563} (\bibinfo {year} {1995})},\ \Eprint
  {http://arxiv.org/abs/hep-ph/9502366} {arXiv:hep-ph/9502366 [hep-ph]}
  \BibitemShut {NoStop}%
\bibitem [{\citenamefont {Bernard}\ \emph {et~al.}(1995)\citenamefont
  {Bernard}, \citenamefont {Kaiser},\ and\ \citenamefont
  {Meissner}}]{Bernard:1995dp}%
  \BibitemOpen
  \bibfield  {author} {\bibinfo {author} {\bibfnamefont {V.}~\bibnamefont
  {Bernard}}, \bibinfo {author} {\bibfnamefont {N.}~\bibnamefont {Kaiser}}, \
  and\ \bibinfo {author} {\bibfnamefont {U.-G.}\ \bibnamefont {Meissner}},\
  }\href {\doibase 10.1142/S0218301395000092} {\bibfield  {journal} {\bibinfo
  {journal} {Int. J. Mod. Phys.}\ }\textbf {\bibinfo {volume} {E4}},\ \bibinfo
  {pages} {193} (\bibinfo {year} {1995})},\ \Eprint
  {http://arxiv.org/abs/hep-ph/9501384} {arXiv:hep-ph/9501384 [hep-ph]}
  \BibitemShut {NoStop}%
\bibitem [{\citenamefont {Beane}\ \emph {et~al.}(2001)\citenamefont {Beane},
  \citenamefont {Bedaque}, \citenamefont {Childress}, \citenamefont
  {Kryjevski}, \citenamefont {McGuire},\ and\ \citenamefont {van
  Kolck}}]{Beane:2000wh}%
  \BibitemOpen
  \bibfield  {author} {\bibinfo {author} {\bibfnamefont {S.~R.}\ \bibnamefont
  {Beane}}, \bibinfo {author} {\bibfnamefont {P.~F.}\ \bibnamefont {Bedaque}},
  \bibinfo {author} {\bibfnamefont {L.}~\bibnamefont {Childress}}, \bibinfo
  {author} {\bibfnamefont {A.}~\bibnamefont {Kryjevski}}, \bibinfo {author}
  {\bibfnamefont {J.}~\bibnamefont {McGuire}}, \ and\ \bibinfo {author}
  {\bibfnamefont {U.}~\bibnamefont {van Kolck}},\ }\href {\doibase
  10.1103/PhysRevA.64.042103} {\bibfield  {journal} {\bibinfo  {journal} {Phys.
  Rev.}\ }\textbf {\bibinfo {volume} {A64}},\ \bibinfo {pages} {042103}
  (\bibinfo {year} {2001})},\ \Eprint {http://arxiv.org/abs/quant-ph/0010073}
  {arXiv:quant-ph/0010073 [quant-ph]} \BibitemShut {NoStop}%
\bibitem [{\citenamefont {Nogga}\ \emph {et~al.}(2005)\citenamefont {Nogga},
  \citenamefont {Timmermans},\ and\ \citenamefont {van Kolck}}]{Nogga:2005hy}%
  \BibitemOpen
  \bibfield  {author} {\bibinfo {author} {\bibfnamefont {A.}~\bibnamefont
  {Nogga}}, \bibinfo {author} {\bibfnamefont {R.~G.~E.}\ \bibnamefont
  {Timmermans}}, \ and\ \bibinfo {author} {\bibfnamefont {U.}~\bibnamefont {van
  Kolck}},\ }\href {\doibase 10.1103/PhysRevC.72.054006} {\bibfield  {journal}
  {\bibinfo  {journal} {Phys. Rev.}\ }\textbf {\bibinfo {volume} {C72}},\
  \bibinfo {pages} {054006} (\bibinfo {year} {2005})},\ \Eprint
  {http://arxiv.org/abs/nucl-th/0506005} {arXiv:nucl-th/0506005 [nucl-th]}
  \BibitemShut {NoStop}%
\bibitem [{\citenamefont {Long}\ and\ \citenamefont {van
  Kolck}(2008)}]{Long:2007vp}%
  \BibitemOpen
  \bibfield  {author} {\bibinfo {author} {\bibfnamefont {B.}~\bibnamefont
  {Long}}\ and\ \bibinfo {author} {\bibfnamefont {U.}~\bibnamefont {van
  Kolck}},\ }\href {\doibase 10.1016/j.aop.2008.01.003} {\bibfield  {journal}
  {\bibinfo  {journal} {Annals Phys.}\ }\textbf {\bibinfo {volume} {323}},\
  \bibinfo {pages} {1304} (\bibinfo {year} {2008})},\ \Eprint
  {http://arxiv.org/abs/0707.4325} {arXiv:0707.4325 [quant-ph]} \BibitemShut
  {NoStop}%
\bibitem [{\citenamefont {Long}\ and\ \citenamefont
  {Yang}(2011)}]{Long:2011qx}%
  \BibitemOpen
  \bibfield  {author} {\bibinfo {author} {\bibfnamefont {B.}~\bibnamefont
  {Long}}\ and\ \bibinfo {author} {\bibfnamefont {C.~J.}\ \bibnamefont
  {Yang}},\ }\href {\doibase 10.1103/PhysRevC.84.057001} {\bibfield  {journal}
  {\bibinfo  {journal} {Phys. Rev.}\ }\textbf {\bibinfo {volume} {C84}},\
  \bibinfo {pages} {057001} (\bibinfo {year} {2011})},\ \Eprint
  {http://arxiv.org/abs/1108.0985} {arXiv:1108.0985 [nucl-th]} \BibitemShut
  {NoStop}%
\bibitem [{\citenamefont {Long}\ and\ \citenamefont
  {Yang}(2012{\natexlab{a}})}]{Long:2011xw}%
  \BibitemOpen
  \bibfield  {author} {\bibinfo {author} {\bibfnamefont {B.}~\bibnamefont
  {Long}}\ and\ \bibinfo {author} {\bibfnamefont {C.~J.}\ \bibnamefont
  {Yang}},\ }\href {\doibase 10.1103/PhysRevC.85.034002} {\bibfield  {journal}
  {\bibinfo  {journal} {Phys. Rev.}\ }\textbf {\bibinfo {volume} {C85}},\
  \bibinfo {pages} {034002} (\bibinfo {year} {2012}{\natexlab{a}})},\ \Eprint
  {http://arxiv.org/abs/1111.3993} {arXiv:1111.3993 [nucl-th]} \BibitemShut
  {NoStop}%
\bibitem [{\citenamefont {Long}\ and\ \citenamefont
  {Yang}(2012{\natexlab{b}})}]{Long:2012ve}%
  \BibitemOpen
  \bibfield  {author} {\bibinfo {author} {\bibfnamefont {B.}~\bibnamefont
  {Long}}\ and\ \bibinfo {author} {\bibfnamefont {C.~J.}\ \bibnamefont
  {Yang}},\ }\href {\doibase 10.1103/PhysRevC.86.024001} {\bibfield  {journal}
  {\bibinfo  {journal} {Phys. Rev.}\ }\textbf {\bibinfo {volume} {C86}},\
  \bibinfo {pages} {024001} (\bibinfo {year} {2012}{\natexlab{b}})},\ \Eprint
  {http://arxiv.org/abs/1202.4053} {arXiv:1202.4053 [nucl-th]} \BibitemShut
  {NoStop}%
\bibitem [{\citenamefont {Birse}(2006)}]{Birse:2005um}%
  \BibitemOpen
  \bibfield  {author} {\bibinfo {author} {\bibfnamefont {M.~C.}\ \bibnamefont
  {Birse}},\ }\href {\doibase 10.1103/PhysRevC.74.014003} {\bibfield  {journal}
  {\bibinfo  {journal} {Phys. Rev.}\ }\textbf {\bibinfo {volume} {C74}},\
  \bibinfo {pages} {014003} (\bibinfo {year} {2006})},\ \Eprint
  {http://arxiv.org/abs/nucl-th/0507077} {arXiv:nucl-th/0507077 [nucl-th]}
  \BibitemShut {NoStop}%
\bibitem [{\citenamefont {Barford}\ and\ \citenamefont
  {Birse}(2003)}]{Barford:2002je}%
  \BibitemOpen
  \bibfield  {author} {\bibinfo {author} {\bibfnamefont {T.}~\bibnamefont
  {Barford}}\ and\ \bibinfo {author} {\bibfnamefont {M.~C.}\ \bibnamefont
  {Birse}},\ }\href {\doibase 10.1103/PhysRevC.67.064006} {\bibfield  {journal}
  {\bibinfo  {journal} {Phys. Rev.}\ }\textbf {\bibinfo {volume} {C67}},\
  \bibinfo {pages} {064006} (\bibinfo {year} {2003})},\ \Eprint
  {http://arxiv.org/abs/hep-ph/0206146} {arXiv:hep-ph/0206146 [hep-ph]}
  \BibitemShut {NoStop}%
\bibitem [{\citenamefont {Pavón~Valderrama}\ and\ \citenamefont
  {Phillips}(2015)}]{Valderrama:2014vra}%
  \BibitemOpen
  \bibfield  {author} {\bibinfo {author} {\bibfnamefont {M.}~\bibnamefont
  {Pavón~Valderrama}}\ and\ \bibinfo {author} {\bibfnamefont {D.~R.}\
  \bibnamefont {Phillips}},\ }\href {\doibase 10.1103/PhysRevLett.114.082502}
  {\bibfield  {journal} {\bibinfo  {journal} {Phys. Rev. Lett.}\ }\textbf
  {\bibinfo {volume} {114}},\ \bibinfo {pages} {082502} (\bibinfo {year}
  {2015})},\ \Eprint {http://arxiv.org/abs/1407.0437} {arXiv:1407.0437
  [nucl-th]} \BibitemShut {NoStop}%
\bibitem [{\citenamefont {Epelbaum}\ and\ \citenamefont
  {Gegelia}(2012)}]{Epelbaum:2012ua}%
  \BibitemOpen
  \bibfield  {author} {\bibinfo {author} {\bibfnamefont {E.}~\bibnamefont
  {Epelbaum}}\ and\ \bibinfo {author} {\bibfnamefont {J.}~\bibnamefont
  {Gegelia}},\ }\href {\doibase 10.1016/j.physletb.2012.08.025} {\bibfield
  {journal} {\bibinfo  {journal} {Phys. Lett.}\ }\textbf {\bibinfo {volume}
  {B716}},\ \bibinfo {pages} {338} (\bibinfo {year} {2012})},\ \Eprint
  {http://arxiv.org/abs/1207.2420} {arXiv:1207.2420 [nucl-th]} \BibitemShut
  {NoStop}%
\bibitem [{\citenamefont {Li}\ \emph {et~al.}(2016)\citenamefont {Li},
  \citenamefont {Ren}, \citenamefont {Geng},\ and\ \citenamefont
  {Long}}]{Li:2016paq}%
  \BibitemOpen
  \bibfield  {author} {\bibinfo {author} {\bibfnamefont {K.-W.}\ \bibnamefont
  {Li}}, \bibinfo {author} {\bibfnamefont {X.-L.}\ \bibnamefont {Ren}},
  \bibinfo {author} {\bibfnamefont {L.-S.}\ \bibnamefont {Geng}}, \ and\
  \bibinfo {author} {\bibfnamefont {B.}~\bibnamefont {Long}},\ }\href {\doibase
  10.1103/PhysRevD.94.014029} {\bibfield  {journal} {\bibinfo  {journal} {Phys.
  Rev.}\ }\textbf {\bibinfo {volume} {D94}},\ \bibinfo {pages} {014029}
  (\bibinfo {year} {2016})},\ \Eprint {http://arxiv.org/abs/1603.07802}
  {arXiv:1603.07802 [hep-ph]} \BibitemShut {NoStop}%
\bibitem [{\citenamefont {Baru}\ \emph {et~al.}(2019)\citenamefont {Baru},
  \citenamefont {Epelbaum}, \citenamefont {Gegelia},\ and\ \citenamefont
  {Ren}}]{Baru:2019ndr}%
  \BibitemOpen
  \bibfield  {author} {\bibinfo {author} {\bibfnamefont {V.}~\bibnamefont
  {Baru}}, \bibinfo {author} {\bibfnamefont {E.}~\bibnamefont {Epelbaum}},
  \bibinfo {author} {\bibfnamefont {J.}~\bibnamefont {Gegelia}}, \ and\
  \bibinfo {author} {\bibfnamefont {X.~L.}\ \bibnamefont {Ren}},\ }\href
  {\doibase 10.1016/j.physletb.2019.134987} {\bibfield  {journal} {\bibinfo
  {journal} {Phys. Lett.}\ }\textbf {\bibinfo {volume} {B798}},\ \bibinfo
  {pages} {134987} (\bibinfo {year} {2019})},\ \Eprint
  {http://arxiv.org/abs/1905.02116} {arXiv:1905.02116 [nucl-th]} \BibitemShut
  {NoStop}%
\bibitem [{\citenamefont {Ren}\ \emph {et~al.}({\natexlab{a}})\citenamefont
  {Ren}, \citenamefont {Epelbaum},\ and\ \citenamefont
  {Gegelia}}]{Ren:2019qow}%
  \BibitemOpen
  \bibfield  {author} {\bibinfo {author} {\bibfnamefont {X.~L.}\ \bibnamefont
  {Ren}}, \bibinfo {author} {\bibfnamefont {E.}~\bibnamefont {Epelbaum}}, \
  and\ \bibinfo {author} {\bibfnamefont {J.}~\bibnamefont {Gegelia}},\
  }\href@noop {} {\  ({\natexlab{a}})},\ \Eprint
  {http://arxiv.org/abs/1911.05616} {arXiv:1911.05616 [nucl-th]} \BibitemShut
  {NoStop}%
\bibitem [{\citenamefont {Ren}\ \emph {et~al.}(2018)\citenamefont {Ren},
  \citenamefont {Li}, \citenamefont {Geng}, \citenamefont {Long}, \citenamefont
  {Ring},\ and\ \citenamefont {Meng}}]{Ren:2016jna}%
  \BibitemOpen
  \bibfield  {author} {\bibinfo {author} {\bibfnamefont {X.-L.}\ \bibnamefont
  {Ren}}, \bibinfo {author} {\bibfnamefont {K.-W.}\ \bibnamefont {Li}},
  \bibinfo {author} {\bibfnamefont {L.-S.}\ \bibnamefont {Geng}}, \bibinfo
  {author} {\bibfnamefont {B.-W.}\ \bibnamefont {Long}}, \bibinfo {author}
  {\bibfnamefont {P.}~\bibnamefont {Ring}}, \ and\ \bibinfo {author}
  {\bibfnamefont {J.}~\bibnamefont {Meng}},\ }\href {\doibase
  10.1088/1674-1137/42/1/014103} {\bibfield  {journal} {\bibinfo  {journal}
  {Chin. Phys.}\ }\textbf {\bibinfo {volume} {C42}},\ \bibinfo {pages} {014103}
  (\bibinfo {year} {2018})},\ \Eprint {http://arxiv.org/abs/1611.08475}
  {arXiv:1611.08475 [nucl-th]} \BibitemShut {NoStop}%
\bibitem [{\citenamefont {Kadyshevsky}(1968)}]{Kadyshevsky:1967rs}%
  \BibitemOpen
  \bibfield  {author} {\bibinfo {author} {\bibfnamefont {V.~G.}\ \bibnamefont
  {Kadyshevsky}},\ }\href {\doibase 10.1016/0550-3213(68)90274-5} {\bibfield
  {journal} {\bibinfo  {journal} {Nucl. Phys.}\ }\textbf {\bibinfo {volume}
  {B6}},\ \bibinfo {pages} {125} (\bibinfo {year} {1968})}\BibitemShut
  {NoStop}%
\bibitem [{\citenamefont {Blankenbecler}\ and\ \citenamefont
  {Sugar}(1966)}]{Blankenbecler:1965gx}%
  \BibitemOpen
  \bibfield  {author} {\bibinfo {author} {\bibfnamefont {R.}~\bibnamefont
  {Blankenbecler}}\ and\ \bibinfo {author} {\bibfnamefont {R.}~\bibnamefont
  {Sugar}},\ }\href {\doibase 10.1103/PhysRev.142.1051} {\bibfield  {journal}
  {\bibinfo  {journal} {Phys. Rev.}\ }\textbf {\bibinfo {volume} {142}},\
  \bibinfo {pages} {1051} (\bibinfo {year} {1966})}\BibitemShut {NoStop}%
\bibitem [{\citenamefont {Li}\ \emph {et~al.}(2018{\natexlab{a}})\citenamefont
  {Li}, \citenamefont {Ren}, \citenamefont {Geng},\ and\ \citenamefont
  {Long}}]{Li:2016mln}%
  \BibitemOpen
  \bibfield  {author} {\bibinfo {author} {\bibfnamefont {K.-W.}\ \bibnamefont
  {Li}}, \bibinfo {author} {\bibfnamefont {X.-L.}\ \bibnamefont {Ren}},
  \bibinfo {author} {\bibfnamefont {L.-S.}\ \bibnamefont {Geng}}, \ and\
  \bibinfo {author} {\bibfnamefont {B.-W.}\ \bibnamefont {Long}},\ }\href
  {\doibase 10.1088/1674-1137/42/1/014105} {\bibfield  {journal} {\bibinfo
  {journal} {Chin. Phys.}\ }\textbf {\bibinfo {volume} {C42}},\ \bibinfo
  {pages} {014105} (\bibinfo {year} {2018}{\natexlab{a}})},\ \Eprint
  {http://arxiv.org/abs/1612.08482} {arXiv:1612.08482 [nucl-th]} \BibitemShut
  {NoStop}%
\bibitem [{\citenamefont {Song}\ \emph {et~al.}(2018)\citenamefont {Song},
  \citenamefont {Li},\ and\ \citenamefont {Geng}}]{Song:2018qqm}%
  \BibitemOpen
  \bibfield  {author} {\bibinfo {author} {\bibfnamefont {J.}~\bibnamefont
  {Song}}, \bibinfo {author} {\bibfnamefont {K.-W.}\ \bibnamefont {Li}}, \ and\
  \bibinfo {author} {\bibfnamefont {L.-S.}\ \bibnamefont {Geng}},\ }\href
  {\doibase 10.1103/PhysRevC.97.065201} {\bibfield  {journal} {\bibinfo
  {journal} {Phys. Rev.}\ }\textbf {\bibinfo {volume} {C97}},\ \bibinfo {pages}
  {065201} (\bibinfo {year} {2018})},\ \Eprint
  {http://arxiv.org/abs/1802.04433} {arXiv:1802.04433 [nucl-th]} \BibitemShut
  {NoStop}%
\bibitem [{\citenamefont {Li}\ \emph {et~al.}(2018{\natexlab{b}})\citenamefont
  {Li}, \citenamefont {Hyodo},\ and\ \citenamefont {Geng}}]{Li:2018tbt}%
  \BibitemOpen
  \bibfield  {author} {\bibinfo {author} {\bibfnamefont {K.-W.}\ \bibnamefont
  {Li}}, \bibinfo {author} {\bibfnamefont {T.}~\bibnamefont {Hyodo}}, \ and\
  \bibinfo {author} {\bibfnamefont {L.-S.}\ \bibnamefont {Geng}},\ }\href
  {\doibase 10.1103/PhysRevC.98.065203} {\bibfield  {journal} {\bibinfo
  {journal} {Phys. Rev.}\ }\textbf {\bibinfo {volume} {C98}},\ \bibinfo {pages}
  {065203} (\bibinfo {year} {2018}{\natexlab{b}})},\ \Eprint
  {http://arxiv.org/abs/1809.03199} {arXiv:1809.03199 [nucl-th]} \BibitemShut
  {NoStop}%
\bibitem [{\citenamefont {Ren}\ \emph {et~al.}({\natexlab{b}})\citenamefont
  {Ren}, \citenamefont {Li}, \citenamefont {Geng},\ and\ \citenamefont
  {Meng}}]{Ren:2017yvw}%
  \BibitemOpen
  \bibfield  {author} {\bibinfo {author} {\bibfnamefont {X.-L.}\ \bibnamefont
  {Ren}}, \bibinfo {author} {\bibfnamefont {K.-W.}\ \bibnamefont {Li}},
  \bibinfo {author} {\bibfnamefont {L.-S.}\ \bibnamefont {Geng}}, \ and\
  \bibinfo {author} {\bibfnamefont {J.}~\bibnamefont {Meng}},\ }\href@noop {}
  {\  ({\natexlab{b}})},\ \Eprint {http://arxiv.org/abs/1712.10083}
  {arXiv:1712.10083 [nucl-th]} \BibitemShut {NoStop}%
\bibitem [{\citenamefont {Sánchez~Sánchez}\ \emph {et~al.}(2018)\citenamefont
  {Sánchez~Sánchez}, \citenamefont {Yang}, \citenamefont {Long},\ and\
  \citenamefont {van Kolck}}]{SanchezSanchez:2017tws}%
  \BibitemOpen
  \bibfield  {author} {\bibinfo {author} {\bibfnamefont {M.}~\bibnamefont
  {Sánchez~Sánchez}}, \bibinfo {author} {\bibfnamefont {C.~J.}\ \bibnamefont
  {Yang}}, \bibinfo {author} {\bibfnamefont {B.}~\bibnamefont {Long}}, \ and\
  \bibinfo {author} {\bibfnamefont {U.}~\bibnamefont {van Kolck}},\ }\href
  {\doibase 10.1103/PhysRevC.97.024001} {\bibfield  {journal} {\bibinfo
  {journal} {Phys. Rev.}\ }\textbf {\bibinfo {volume} {C97}},\ \bibinfo {pages}
  {024001} (\bibinfo {year} {2018})},\ \Eprint
  {http://arxiv.org/abs/1704.08524} {arXiv:1704.08524 [nucl-th]} \BibitemShut
  {NoStop}%
\bibitem [{\citenamefont {Gegelia}\ and\ \citenamefont
  {Japaridze}(1999)}]{Gegelia:1999gf}%
  \BibitemOpen
  \bibfield  {author} {\bibinfo {author} {\bibfnamefont {J.}~\bibnamefont
  {Gegelia}}\ and\ \bibinfo {author} {\bibfnamefont {G.}~\bibnamefont
  {Japaridze}},\ }\href {\doibase 10.1103/PhysRevD.60.114038} {\bibfield
  {journal} {\bibinfo  {journal} {Phys. Rev.}\ }\textbf {\bibinfo {volume}
  {D60}},\ \bibinfo {pages} {114038} (\bibinfo {year} {1999})},\ \Eprint
  {http://arxiv.org/abs/hep-ph/9908377} {arXiv:hep-ph/9908377 [hep-ph]}
  \BibitemShut {NoStop}%
\bibitem [{\citenamefont {Fuchs}\ \emph {et~al.}(2003)\citenamefont {Fuchs},
  \citenamefont {Gegelia}, \citenamefont {Japaridze},\ and\ \citenamefont
  {Scherer}}]{Fuchs:2003qc}%
  \BibitemOpen
  \bibfield  {author} {\bibinfo {author} {\bibfnamefont {T.}~\bibnamefont
  {Fuchs}}, \bibinfo {author} {\bibfnamefont {J.}~\bibnamefont {Gegelia}},
  \bibinfo {author} {\bibfnamefont {G.}~\bibnamefont {Japaridze}}, \ and\
  \bibinfo {author} {\bibfnamefont {S.}~\bibnamefont {Scherer}},\ }\href
  {\doibase 10.1103/PhysRevD.68.056005} {\bibfield  {journal} {\bibinfo
  {journal} {Phys. Rev.}\ }\textbf {\bibinfo {volume} {D68}},\ \bibinfo {pages}
  {056005} (\bibinfo {year} {2003})},\ \Eprint
  {http://arxiv.org/abs/hep-ph/0302117} {arXiv:hep-ph/0302117 [hep-ph]}
  \BibitemShut {NoStop}%
\bibitem [{\citenamefont {Geng}(2013)}]{Geng:2013xn}%
  \BibitemOpen
  \bibfield  {author} {\bibinfo {author} {\bibfnamefont {L.}~\bibnamefont
  {Geng}},\ }\href {\doibase 10.1007/s11467-013-0327-7} {\bibfield  {journal}
  {\bibinfo  {journal} {Front. Phys.(Beijing)}\ }\textbf {\bibinfo {volume}
  {8}},\ \bibinfo {pages} {328} (\bibinfo {year} {2013})},\ \Eprint
  {http://arxiv.org/abs/1301.6815} {arXiv:1301.6815 [nucl-th]} \BibitemShut
  {NoStop}%
\bibitem [{\citenamefont {Soto}\ and\ \citenamefont
  {Tarrus}(2008)}]{Soto:2007pg}%
  \BibitemOpen
  \bibfield  {author} {\bibinfo {author} {\bibfnamefont {J.}~\bibnamefont
  {Soto}}\ and\ \bibinfo {author} {\bibfnamefont {J.}~\bibnamefont {Tarrus}},\
  }\href {\doibase 10.1103/PhysRevC.78.024003} {\bibfield  {journal} {\bibinfo
  {journal} {Phys. Rev.}\ }\textbf {\bibinfo {volume} {C78}},\ \bibinfo {pages}
  {024003} (\bibinfo {year} {2008})},\ \Eprint {http://arxiv.org/abs/0712.3404}
  {arXiv:0712.3404 [nucl-th]} \BibitemShut {NoStop}%
\bibitem [{\citenamefont {Long}(2013)}]{Long:2013cya}%
  \BibitemOpen
  \bibfield  {author} {\bibinfo {author} {\bibfnamefont {B.}~\bibnamefont
  {Long}},\ }\href {\doibase 10.1103/PhysRevC.88.014002} {\bibfield  {journal}
  {\bibinfo  {journal} {Phys. Rev.}\ }\textbf {\bibinfo {volume} {C88}},\
  \bibinfo {pages} {014002} (\bibinfo {year} {2013})},\ \Eprint
  {http://arxiv.org/abs/1304.7382} {arXiv:1304.7382 [nucl-th]} \BibitemShut
  {NoStop}%
\bibitem [{\citenamefont {Xiao}\ \emph {et~al.}(2019)\citenamefont {Xiao},
  \citenamefont {Geng},\ and\ \citenamefont {Ren}}]{Xiao:2018jot}%
  \BibitemOpen
  \bibfield  {author} {\bibinfo {author} {\bibfnamefont {Y.}~\bibnamefont
  {Xiao}}, \bibinfo {author} {\bibfnamefont {L.-S.}\ \bibnamefont {Geng}}, \
  and\ \bibinfo {author} {\bibfnamefont {X.-L.}\ \bibnamefont {Ren}},\ }\href
  {\doibase 10.1103/PhysRevC.99.024004} {\bibfield  {journal} {\bibinfo
  {journal} {Phys. Rev.}\ }\textbf {\bibinfo {volume} {C99}},\ \bibinfo {pages}
  {024004} (\bibinfo {year} {2019})},\ \Eprint
  {http://arxiv.org/abs/1812.03005} {arXiv:1812.03005 [nucl-th]} \BibitemShut
  {NoStop}%
\bibitem [{\citenamefont {Stoks}\ \emph {et~al.}(1993)\citenamefont {Stoks},
  \citenamefont {Klomp}, \citenamefont {Rentmeester},\ and\ \citenamefont
  {de~Swart}}]{Stoks:1993tb}%
  \BibitemOpen
  \bibfield  {author} {\bibinfo {author} {\bibfnamefont {V.~G.~J.}\
  \bibnamefont {Stoks}}, \bibinfo {author} {\bibfnamefont {R.~A.~M.}\
  \bibnamefont {Klomp}}, \bibinfo {author} {\bibfnamefont {M.~C.~M.}\
  \bibnamefont {Rentmeester}}, \ and\ \bibinfo {author} {\bibfnamefont {J.~J.}\
  \bibnamefont {de~Swart}},\ }\href {\doibase 10.1103/PhysRevC.48.792}
  {\bibfield  {journal} {\bibinfo  {journal} {Phys. Rev.}\ }\textbf {\bibinfo
  {volume} {C48}},\ \bibinfo {pages} {792} (\bibinfo {year}
  {1993})}\BibitemShut {NoStop}%
\bibitem [{\citenamefont {Yang}\ \emph
  {et~al.}(2009{\natexlab{a}})\citenamefont {Yang}, \citenamefont {Elster},\
  and\ \citenamefont {Phillips}}]{Yang:2009kx}%
  \BibitemOpen
  \bibfield  {author} {\bibinfo {author} {\bibfnamefont {C.~J.}\ \bibnamefont
  {Yang}}, \bibinfo {author} {\bibfnamefont {C.}~\bibnamefont {Elster}}, \ and\
  \bibinfo {author} {\bibfnamefont {D.~R.}\ \bibnamefont {Phillips}},\ }\href
  {\doibase 10.1103/PhysRevC.80.034002} {\bibfield  {journal} {\bibinfo
  {journal} {Phys. Rev.}\ }\textbf {\bibinfo {volume} {C80}},\ \bibinfo {pages}
  {034002} (\bibinfo {year} {2009}{\natexlab{a}})},\ \Eprint
  {http://arxiv.org/abs/0901.2663} {arXiv:0901.2663 [nucl-th]} \BibitemShut
  {NoStop}%
\bibitem [{\citenamefont {Wigner}(1955)}]{Wigner:1955zz}%
  \BibitemOpen
  \bibfield  {author} {\bibinfo {author} {\bibfnamefont {E.~P.}\ \bibnamefont
  {Wigner}},\ }\href {\doibase 10.1103/PhysRev.98.145} {\bibfield  {journal}
  {\bibinfo  {journal} {Phys. Rev.}\ }\textbf {\bibinfo {volume} {98}},\
  \bibinfo {pages} {145} (\bibinfo {year} {1955})}\BibitemShut {NoStop}%
\bibitem [{\citenamefont {Phillips}\ and\ \citenamefont
  {Cohen}(1997)}]{Phillips:1996ae}%
  \BibitemOpen
  \bibfield  {author} {\bibinfo {author} {\bibfnamefont {D.~R.}\ \bibnamefont
  {Phillips}}\ and\ \bibinfo {author} {\bibfnamefont {T.~D.}\ \bibnamefont
  {Cohen}},\ }\href {\doibase 10.1016/S0370-2693(96)01411-6} {\bibfield
  {journal} {\bibinfo  {journal} {Phys. Lett.}\ }\textbf {\bibinfo {volume}
  {B390}},\ \bibinfo {pages} {7} (\bibinfo {year} {1997})},\ \Eprint
  {http://arxiv.org/abs/nucl-th/9607048} {arXiv:nucl-th/9607048 [nucl-th]}
  \BibitemShut {NoStop}%
\bibitem [{\citenamefont {Phillips}\ \emph {et~al.}(1998)\citenamefont
  {Phillips}, \citenamefont {Beane},\ and\ \citenamefont
  {Cohen}}]{Phillips:1997xu}%
  \BibitemOpen
  \bibfield  {author} {\bibinfo {author} {\bibfnamefont {D.~R.}\ \bibnamefont
  {Phillips}}, \bibinfo {author} {\bibfnamefont {S.~R.}\ \bibnamefont {Beane}},
  \ and\ \bibinfo {author} {\bibfnamefont {T.~D.}\ \bibnamefont {Cohen}},\
  }\href {\doibase 10.1006/aphy.1997.5771} {\bibfield  {journal} {\bibinfo
  {journal} {Annals Phys.}\ }\textbf {\bibinfo {volume} {263}},\ \bibinfo
  {pages} {255} (\bibinfo {year} {1998})},\ \Eprint
  {http://arxiv.org/abs/hep-th/9706070} {arXiv:hep-th/9706070 [hep-th]}
  \BibitemShut {NoStop}%
\bibitem [{\citenamefont {Wu}\ and\ \citenamefont {Long}(2019)}]{Wu:2018lai}%
  \BibitemOpen
  \bibfield  {author} {\bibinfo {author} {\bibfnamefont {S.}~\bibnamefont
  {Wu}}\ and\ \bibinfo {author} {\bibfnamefont {B.}~\bibnamefont {Long}},\
  }\href {\doibase 10.1103/PhysRevC.99.024003} {\bibfield  {journal} {\bibinfo
  {journal} {Phys. Rev.}\ }\textbf {\bibinfo {volume} {C99}},\ \bibinfo {pages}
  {024003} (\bibinfo {year} {2019})},\ \Eprint
  {http://arxiv.org/abs/1807.04407} {arXiv:1807.04407 [nucl-th]} \BibitemShut
  {NoStop}%
\bibitem [{\citenamefont {Kaplan}()}]{Kaplan:2019znu}%
  \BibitemOpen
  \bibfield  {author} {\bibinfo {author} {\bibfnamefont {D.~B.}\ \bibnamefont
  {Kaplan}},\ }\href@noop {} {\ }\Eprint {http://arxiv.org/abs/1905.07485}
  {arXiv:1905.07485 [nucl-th]} \BibitemShut {NoStop}%
\bibitem [{\citenamefont {Yang}\ \emph
  {et~al.}(2009{\natexlab{b}})\citenamefont {Yang}, \citenamefont {Elster},\
  and\ \citenamefont {Phillips}}]{Yang:2009pn}%
  \BibitemOpen
  \bibfield  {author} {\bibinfo {author} {\bibfnamefont {C.~J.}\ \bibnamefont
  {Yang}}, \bibinfo {author} {\bibfnamefont {C.}~\bibnamefont {Elster}}, \ and\
  \bibinfo {author} {\bibfnamefont {D.~R.}\ \bibnamefont {Phillips}},\ }\href
  {\doibase 10.1103/PhysRevC.80.044002} {\bibfield  {journal} {\bibinfo
  {journal} {Phys. Rev.}\ }\textbf {\bibinfo {volume} {C80}},\ \bibinfo {pages}
  {044002} (\bibinfo {year} {2009}{\natexlab{b}})},\ \Eprint
  {http://arxiv.org/abs/0905.4943} {arXiv:0905.4943 [nucl-th]} \BibitemShut
  {NoStop}%
\end{thebibliography}%

\end{document}